\documentclass[aps,prc,superscriptaddress,floats,floatfix,reprint,reprintnumbers]{revtex4}
\usepackage{blindtext}
\pdfoutput=1
\usepackage{epsfig}
\usepackage{graphicx}% Include figure files
\usepackage{dcolumn}% Align table columns on decimal point
\usepackage{bm}% bold math
\usepackage{booktabs}
\usepackage{color}
\usepackage{mathrsfs}
\usepackage{ulem}
\usepackage{epstopdf}
\graphicspath{{plots/}}%directory of my plots
\usepackage{verbatim}
\usepackage{indentfirst}
\usepackage{float}
\usepackage[colorlinks,
           bookmarks=true,
           linkcolor=blue,
           urlcolor=blue,
            anchorcolor=black,
            citecolor=blue]{hyperref}
\usepackage[bf,raggedright,indentafter]{titlesec}
\usepackage{rotating}
\usepackage{booktabs}
\usepackage{tabularx}
\usepackage{makecell}
\usepackage{hypcap}
\usepackage{subfigure}%
\usepackage{tabularx}

\usepackage{array}

\begin{document}
\title{\Large Studying Radiative Baryon Decays with the SU(3) Flavor Symmetry }
\author{Ru-Min Wang$^{1,\dagger}$,~~Xiao-Dong Cheng$^{2,\S}$,~~Ying-Ying Fan$^{2,\diamondsuit}$,~~Jie-Lei Zhang$^{2,\sharp}$,~~Yuan-Guo Xu$^{1,\ddag}$\\
$^1${\scriptsize College of Physics and Communication Electronics, JiangXi Normal University, NanChang, JiangXi 330022, China}\\
$^2${\scriptsize College of Physics and Electronic Engineering, XinYang Normal University, XinYang, Henan 464000, China}\\
$^\dagger${\scriptsize ruminwang@sina.com}~~~$^\S${\scriptsize chengxd@mails.ccnu.edu.cn}~~~$^\diamondsuit${\scriptsize fyy163@126.com}~~~
$^\sharp${\scriptsize zhangjielei@ihep.ac.cn}~~~$^\ddag${\scriptsize yuanguoxv@163.com}}

\begin{abstract}
The weak and electromagnetic radiative baryon decays of
octet $T_{8}$,  decuplet $T_{10}$,  single charmed anti-triplet $T_{c3}$ and sextet $T_{c6}$,  single heavy bottomed anti-triplet $T_{b3}$ and  sextet $T_{b6}$ are investigated by using  SU(3) flavor symmetry irreducible representation approach.
 We analyze the contributions from a single quark transition $q_1\to q_2\gamma$ and $W$ exchange  transitions,  and find that the amplitudes  could be easily related by SU(3)  flavor symmetry in  the $T_{b3,b6}$ weak radiative decays,  $T_{c3,c6}$ weak radiative decays,  $T_{10}\to T_{8}\gamma $ weak decays,   $T_{10}\to T'_{10}\gamma $ weak decays and   $T_{10}\to T_{8}\gamma $ electromagnetic decays. Nevertheless, the amplitude relations are  a little complex in the $T_{8}\to T'_{8}\gamma$ and $T_{8}\to T_{10}\gamma$ weak decays due to quark antisymmetry in $T_{8}$ and $W$ exchange contributions. Predictions for  branching ratios of $\Lambda^{0}_b\to n\gamma$, $\Xi^{-}_b\to \Xi^-\gamma$, $\Xi^{-}_b\to \Sigma^-\gamma$, $\Xi^{0}_b\to \Sigma^0\gamma$, $\Xi^{0}_b\to \Lambda^0\gamma$, $\Xi^{0}_b\to \Xi^0\gamma$, $\Xi^{*}\to \Xi\gamma$, $\Sigma^{*0}\to \Sigma^{0}\gamma$, $\Delta^0\to n\gamma$ and $\Delta^+\to p\gamma$ are given. The results in this work can be used to test SU(3) flavor symmetry approach in the radiative baryon decays by the future experiments at BESIII, LHCb and Belle-II.

\end{abstract}

\maketitle

%\newpage
\section{INTRODUCTION}
Radiative weak decays have attracted a lot of attention for a long time in both
theory and  experiment,  since they could give us a chance to study the interplay of the electromagnetic, weak and strong interactions, to test the standard model  and to probe new physics. A large number of bottomed baryons, charmed baryon and hyperons are produced at the LHC \cite{Cerri:2018ypt,Aaij:2017ddf,Junior:2018odx},
 significant experimental
progresses about $\Lambda_b^0$ baryon rare decays have been
achieved recently at LHCb, and one of them is that  radiative decay $\Lambda^0_b\to\Lambda^0\gamma$ has been observed with a branching ratio of $(7.1\pm1.5\pm0.6\pm0.7)\times10^{-6}$ for the first time  \cite{Aaij:2019hhx}.
Furthermore, many radiative weak decays of strange  baryons have been measured \cite{PDG2020}, and there are longstanding theoretical difficulties to explain the experimental data  \cite{Lach:1995we,Donoghue:1985rk}.   Now the sensitivity for measurements of hyperon decays  is in the range of $10^{-5}-10^{-8}$ at the BESIII \cite{Li:2016tlt,Bigi:2017eni,Asner:2008nq,Ablikim:2018zay}.
Therefore, more baryon radiative decays  will be detected by the experiments in the near future, so  it's feasible to explore these decays now.

Theoretically, due to our poor understanding of QCD at low energy regions, theoretical
calculations of decay amplitudes are not well understood.  SU(3) flavor symmetry has attracted a lot of attentions.  The SU(3) flavor  symmetry approach, which is independent of the detailed dynamics,  offers an opportunity to relate different decay modes.  Nevertheless, it cannot determine
the size of the amplitudes by itself. However, if experimental data are enough, one may use the data to extract the  amplitudes, which can be viewed as predictions based on symmetry.
There are two popular ways of the SU(3) flavor symmetry. One is to construct the
SU(3) irreducible representation amplitude by decomposing effective Hamiltonian. Another way is topological diagram approach, where decay amplitudes are represented by connecting
quark line flows in different ways  and then relate them by the SU(3) symmetry.
The SU(3) irreducible representation approach (IRA) shows a convenient connection with the SU(3) symmetry,  the  topological diagram approach gives a better understanding of dynamics in the different amplitudes.
The SU(3) flavor symmetry  works well  in  bottomed hadron decays \cite{Dery:2020lbc,He:1998rq,He:2000ys,Fu:2003fy,Hsiao:2015iiu,He:2015fwa,He:2015fsa,Deshpande:1994ii,Gronau:1994rj,Gronau:1995hm,Shivashankara:2015cta,Zhou:2016jkv,Cheng:2014rfa,Singer:1995is},  charmed hadron decays \cite{Grossman:2012ry,Pirtskhalava:2011va,Cheng:2012xb,Savage:1989qr,Savage:1991wu,Altarelli:1975ye,Lu:2016ogy,Geng:2017esc,Geng:2018plk,Geng:2017mxn,Geng:2019bfz,Wang:2017azm,Wang:2019dls,Wang:2017gxe,Muller:2015lua} and hyperon decays \cite{Xu:2020jfr,Wang:2019alu,Chang:2014iba,Zenczykowski:2005cs,Zenczykowski:2006se}.

Many weak radiative  decays have been studied  by chiral perturbation theory \cite{Bos:1996ig}, perturbative QCD \cite{He:2006ud},  quark model approach  \cite{Singer:1996xh}, Bethe-Salpeter equation approach \cite{Liu:2019rpm},  relativistic quark model \cite{Faustov:2017ous},  light-cone sum-rule \cite{Aliev:2004ju},  single universal extra dimension scenario \cite{Colangelo:2007jy} and
effective Lagrangian approach \cite{Cheng:1994kp}, etc.
And some electromagnetic radiative baryon decays have been also  studied in Refs. \cite{Ramalho:2020tnn,Junker:2019vvy}.  In this work, we will study the  weak radiative baryon decays with a single quark transitions ($b\to d\gamma$, $b\to s\gamma$,  $c\to u\gamma$, $s\to d\gamma$) and corresponding $W$ exchange transitions  as well as  the electromagnetic radiative decays of $T_{10}\to T_{8}\gamma$  by using the SU(3) IRA.  We  will  firstly construct the
SU(3) irreducible representation amplitudes for different kinds of radiative baryon decays,  secondly  obtain the decay amplitude relations between different decay modes, then use the available data to extract the SU(3) irreducible amplitudes,  and finally  predict the  not-yet-measured modes for further tests in experiments.

This paper is organized as follows. In Sec. II, we will collect the representations for the baryon multiplets and the branching ratio expressions  of the radiative baryon decays. In Sec. III, we will analyze the weak radiative  decays of $T_{b3,b6}$,  $T_{c3,c6}$ and $T_{8,10}$ as well as the   electromagnetic radiative decays $T_{10}\to T_{8}\gamma$.  Our conclusions are given in Sec. IV.

%\newpage
\section{Theoretical Frame for $\mathcal{B}_1\to \mathcal{B}_2\gamma$ }

\subsection{Baryon multiplets}
The light baryons octet $T_{8}$ and decuplet $T_{10}$ under the SU(3) flavor symmetry of $u,d,s$ quarks   can be written as
\begin{eqnarray}
 T_8&=&\left(\begin{array}{ccc}
\frac{\Lambda^0}{\sqrt{6}}+\frac{\Sigma^0}{\sqrt{2}} & \Sigma^+ & p \\
\Sigma^- &\frac{\Lambda^0}{\sqrt{6}}-\frac{\Sigma^0}{\sqrt{2}}  & n \\
\Xi^- & \Xi^0 &-\frac{2\Lambda^0}{\sqrt{6}}
\end{array}\right)\,,\\
%\end{eqnarray}
%\begin{eqnarray}
 T_{10}&=&\frac{1}{\sqrt{3}}\left(
 \left(\begin{array}{ccc}
\sqrt{3}\Delta^{++} & \Delta^{+}  & \Sigma^{*+}  \\
\Delta^{+} & \Delta^{0}  & \frac{\Sigma^{*0}}{\sqrt{2}}  \\
\Sigma^{*+} & \frac{\Sigma^{*0}}{\sqrt{2}}  &\Xi^{*0}
\end{array}\right),~
\left(\begin{array}{ccc}
\Delta^{+} & \Delta^{0}  & \frac{\Sigma^{*0}}{\sqrt{2}}  \\
\Delta^{0} & \sqrt{3}\Delta^{-}  & \Sigma^{*-} \\
\frac{\Sigma^{*0}}{\sqrt{2}} & \Sigma^{*-} &\Xi^{*-}
\end{array}\right),~
\left(\begin{array}{ccc}
\Sigma^{*+} & \frac{\Sigma^{*0}}{\sqrt{2}}  &\Xi^{*0}  \\
\frac{\Sigma^{*0}}{\sqrt{2}} & \Sigma^*{-}  & \Xi^{*-} \\
\Xi^{*0} & \Xi^{*-} &\sqrt{3}\Omega^{-}
\end{array}\right)\right).
\end{eqnarray}
The single charmed anti-triplet $T_{c3}$ and  sextet $T_{c6}$ can be written as
\begin{eqnarray}
T_{c3}&=&(\Xi^0_c,~-\Xi^+_c,~\Lambda^+_c),   ~~~~~~~~~~~~~~~~~~~
T_{c6}=\left(\begin{array}{ccc}
\Sigma^{++}_c & \frac{1}{\sqrt{2}}\Sigma^+_c &  \frac{1}{\sqrt{2}}\Xi^{*+}_c \\
\frac{1}{\sqrt{2}}\Sigma^+_c &\Sigma^{0}_c & \frac{1}{\sqrt{2}}\Xi^{*0}_c \\
 \frac{1}{\sqrt{2}}\Xi^{*+}_c&\frac{1}{\sqrt{2}}\Xi^{*0}_c & \Omega_c
\end{array}\right)\,.
\end{eqnarray}
The  anti-triplet $T_{b3}$ and  sextet $T_{b6}$ with a heavy b quark have a similar form to  $T_{c3}$ and $T_{c6}$, respectively,
\begin{eqnarray}
T_{b3}&=&(\Xi^-_b,~-\Xi^0_b,~\Lambda^0_b),   ~~~~~~~~~~~~~~~~~~~
T_{b6}=\left(\begin{array}{ccc}
\Sigma^{+}_b & \frac{1}{\sqrt{2}}\Sigma^0_b &  \frac{1}{\sqrt{2}}\Xi^{*0}_b \\
\frac{1}{\sqrt{2}}\Sigma^0_b &\Sigma^{-}_b & \frac{1}{\sqrt{2}}\Xi^{*-}_b \\
 \frac{1}{\sqrt{2}}\Xi^{*0}_b&\frac{1}{\sqrt{2}}\Xi^{*-}_b & \Omega_b
\end{array}\right)\,.
\end{eqnarray}

% The doubly heavy baryons can be expressed as the SU(3) triplets
% \begin{eqnarray}
% T_{cc}&=&(\Xi^{++}_{cc},~\Xi^+_{cc},~\Omega^+_{cc}),  ~~~~~~~~~~~~~~~T_{bc}=(\Xi^{+}_{bc},~\Xi^0_{bc},~\Omega^0_{bc}),
% ~~~~~~~~~~~~~~~T_{bb}=(\Xi^{0}_{bb},\Xi^-_{bb},\Omega^-_{bb}).
% \end{eqnarray}

\subsection{Decay branching ratio of  $\mathcal{B}_1\to \mathcal{B}_2\gamma$}

\begin{figure}[b]
\centering
\includegraphics[scale=0.4]{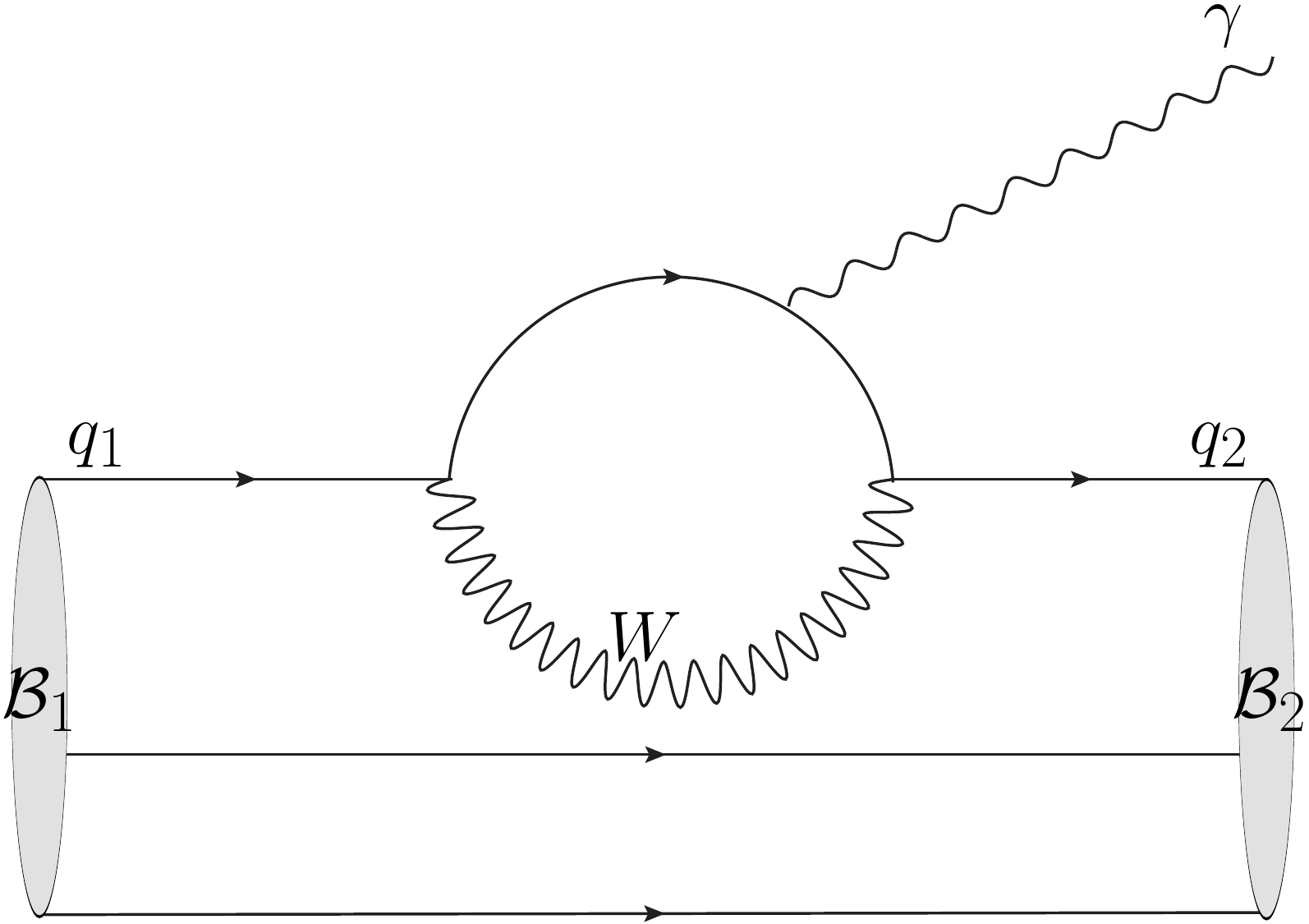}
\caption{Feynman diagram for $\mathcal{B}_1 \to \mathcal{B}_2\gamma$ weak decays via a single quark emission in  the standard model. }
\label{Fig:q1tq2r}
\end{figure}

In the standard model,  the weak radiative baryon decays $\mathcal{B}_1 \to \mathcal{B}_2\gamma$ with  $q_1\to q_2\gamma$ transition  can proceed via loop Feynman diagrams as shown in  Fig. \ref{Fig:q1tq2r}.
The  effective Hamiltonian for $q_1\to q_2\gamma$ transition shown in Fig. \ref{Fig:q1tq2r} can be written as \cite{Buchalla:1995vs}
\begin{eqnarray}
 \mathcal{H}_{eff}(q_1\to q_2\gamma)=-\frac{G_F}{\sqrt{2}}\frac{e\lambda_{q_1q_2}}{4\pi^2}C^{eff}_7
  m_{q_1}\Big(\bar{q}_2i\sigma^{\mu q}P_R q_1 \Big)\epsilon_\mu,\label{EQ:Heff}
 \end{eqnarray}
 where $P_{R}=(1+\gamma_5)/2$,  $\sigma^{\mu q}=\frac{i}{2}(\gamma^\mu\gamma^\nu-\gamma^\nu\gamma^\mu)q_\nu$ with  $q=p_{1}-p_{2} $,   $\epsilon_\mu$ is  the polarization vectors of photon, and the Cabibbo-Kobayashi-Maskawa (CKM) matrix elements $\lambda_{q_1q_2}= V_{tb}V^*_{ts},~V_{tb}V^*_{td},~-V_{us}V^*_{ud}, V^*_{cb}V_{ub}$ for $b\to s \gamma$, $b\to d \gamma$, $s\to d \gamma$, $c\to u \gamma$, respectively.

Then the decay amplitudes can be written as
 \begin{eqnarray}
 \mathcal{M}(\mathcal{B}_1\to \mathcal{B}_2\gamma)&=&\langle \mathcal{B}_2\gamma| \mathcal{H}_{eff}(q_1\to q_2\gamma)|\mathcal{B}_1\rangle\nonumber\\
 &=&-i\frac{G_F}{\sqrt{2}}\frac{e\lambda_{q_1q_2}}{4\pi^2}C^{eff}_7
  m_{q_1}\epsilon_\mu q_\nu\langle \mathcal{B}_2|\bar{q}_2\sigma^{\mu \nu}P_R q_1 |\mathcal{B}_1\rangle.
 \end{eqnarray}
The baryon matrix elements $\langle \mathcal{B}_2|\bar{q}_2\sigma^{\mu \nu}P_R q_1 |\mathcal{B}_1\rangle$ can be parameterized by the form factors,
but not all relevant form factors have been calculated and there is no very reliable method to calculate some form factors at present.
Nevertheless,  the baryon matrix elements  also can be obtained by the SU(3) IRA.  In terms  of the  SU(3) flavor symmetry, baryon states and quark operators can be parameterized into SU(3) tensor forms, while
the polarization vectors $\epsilon_\mu$ are invariant under SU(3) flavor symmetry.  The decay amplitudes in terms of the SU(3) IRA are given in later Tab. \ref{Tab:HTb36tT810}, Tab. \ref{Tab:HTb36tTc36}, Tab. \ref{Tab:HTc36tT810} and Tab. \ref{Tab:HT810tT810W} for $T_{b3,b6}\to T_{8,10}\gamma$, $T_{b3,b6}\to T_{c3,c6}\gamma$, $T_{c3,c6}\to T_{8,10}\gamma $ and  $T_{8,10}\to T^{'}_{8,10}\gamma $  weak decays, respectively.

The branching ratios of  the $\mathcal{B}_1\to \mathcal{B}_2\gamma$ weak decays can be obtained by the decay amplitudes
\begin{eqnarray}
\mathcal{B}(\mathcal{B}_1\to \mathcal{B}_2\gamma)=\frac{\tau_{\mathcal{B}_1}( m^2_{\mathcal{B}_1}-m^2_{\mathcal{B}_2})}{16\pi m^3_{\mathcal{B}_1}}\left|\mathcal{M}(\mathcal{B}_1\to \mathcal{B}_2\gamma) \right|^2.\label{EQ:BrB1tB2rO}
\end{eqnarray}
After extracting  the  masses, the Wilson Coefficients, etc, from $|\mathcal{M}(\mathcal{B}_1\to \mathcal{B}_2\gamma)|^2$,  the branching ratios of  the $T_{b3,c3,8}\to T'_{c3,c6,8,10}\gamma$ and  $T_{b6,c6,10}\to T_{c3,8}\gamma$ weak decays are  \cite{Faustov:2017wbh,Gutsche:2013pp}
\begin{eqnarray}
\mathcal{B}(\mathcal{B}_1\to \mathcal{B}_2\gamma)=\frac{\tau_{\mathcal{B}_1}\alpha_e}{64\pi^4}G^2_Fm_{q_1}^2m^3_{\mathcal{B}_1}|\lambda_{q_1q_2}|^2|C^{eff}_{7\gamma}(m_{q_1})|^2
\left(1-\frac{m^2_{\mathcal{B}_2}}{m^2_{\mathcal{B}_1}}\right)^3\left|A(\mathcal{B}_1\to \mathcal{B}_2\gamma) \right|^2, \label{EQ:BrB1tB2r810}
\end{eqnarray}
where $A(\mathcal{B}_1\to \mathcal{B}_2\gamma)$  may be given by the form factors as
$\left|A(\mathcal{B}_1\to \mathcal{B}_2\gamma) \right|^2=K(|f_2^{TV}(0)|^2+|f_2^{TA}(0)|^2)=K(|h_\bot(0)|^2+|\tilde{h}_\bot(0)|^2)$ with  $K=1$ for the $T_{b3,c3,8}\to T'_{c3,c6,8,10}\gamma$ weak decays and $K=1/2$ for the $T_{b6,c6,10}\to T_{c3,8}\gamma$ weak decays.
As for the $T_{b6,c6}\to T'_{c6,10}\gamma$ weak decays, the branching ratios are \cite{Mannel:2011xg}
\begin{eqnarray}
\mathcal{B}(\mathcal{B}_1\to \mathcal{B}_2\gamma)= \frac{\tau_{\mathcal{B}_1}\alpha_e}{384\pi^4}G^2_F\frac{m_{q_1}^2m^5_{\mathcal{B}_1}}{m^2_{\mathcal{B}_2}}\left(1-\frac{m^2_{\mathcal{B}_2}}{m^2_{\mathcal{B}_1}}\right)^3|\lambda_{q_1q_2}|^2|C^{eff}_{7\gamma}(m_{q_1})|^2
\left|A(\mathcal{B}_1\to \mathcal{B}_2\gamma) \right|^2. \label{EQ:BrB1tB2r10}
\end{eqnarray}
For the electromagnetic  $T_{10}\to T_{8}\gamma$  decays, the expressions of their branching ratios are different from Eq. (\ref{EQ:BrB1tB2r810}), and the following relations will be used to obtain the results \cite{Junker:2019vvy}
 \begin{eqnarray}
\mathcal{B}^{E}(T_{10}\to T_8\gamma)\propto \frac{\tau_{T_{10}}(m^2_{T_{10}}-m^2_{T_{8}})}{m^3_{T_{10}}}\left(m_{T_{10}}-m_{T_{8}}\right)^2\left|A^{E}(T_{10}\to T_{8}\gamma)\right|^2. \label{EQ:BrB1tB2rES}
\end{eqnarray}

\begin{figure}[h]
\centering
\includegraphics[scale=0.96]{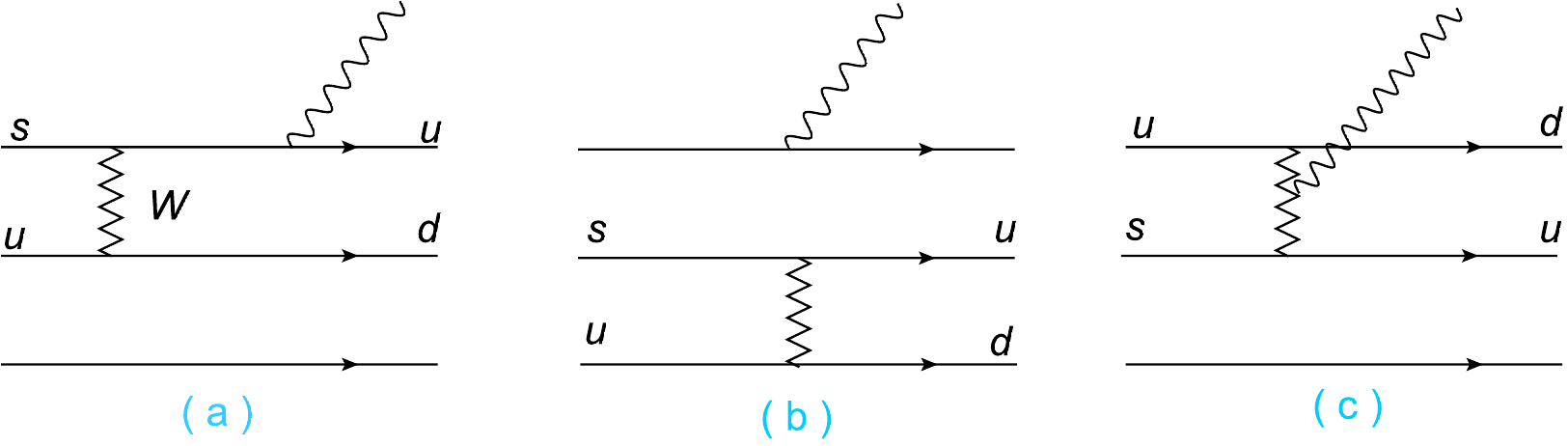}
\caption{Other quark diagrams for weak radiative $\mathcal{B}_1\to \mathcal{B}_2\gamma$ weak decays. (a) two-quark bremsstrahlung, (b) three-quark transition, (c) internal radiation. }
\label{fig:othersdr}
\end{figure}

In addition, according to Refs. \cite{Verma:1988gf,Lach:1995we,Azimov:1996uf}, other three kinds of Feynman diagrams might contribute to the weak baryon  decays. The example for  $s+u\to u+d+\gamma$ is displayed in Fig. \ref{fig:othersdr}. Fig. \ref{fig:othersdr} (a-b) are two-quark and three-quark  transitions with the $W$ exchange, which have been discussed, for examples, in Refs. \cite{Verma:1988gf,Dubovik:2007qg}.
Since Fig. \ref{fig:othersdr} (c) is suppressed by the two $W$ propagators, and its contribution can be safely neglected. We will  consider the W-exchange contributions in Fig. \ref{fig:othersdr} (a-b) in later analysis of SU(3) flavor symmetry.

\section{Results and Analysis}
The theoretical input parameters and the experimental data within the $2\sigma$ errors from Particle Data Group \cite{PDG2020} will be used in our numerical results.

\subsection{$T_{b3,b6}$ weak radiative decays}

The SU(3) flavor structure of the relevant $b\to s,d$ Hamiltonian can been found, for instance,  in Refs. \cite{Zeppenfeld:1980ex,Savage:1989ub,Deshpande:1994ii}.
The  SU(3) IRA decay amplitudes for $T_{b3,b6}\to T_{8,10}\gamma$ decays via $b\to s/d \gamma$  can be parameterized as
\begin{eqnarray}
A(T_{b3} \to T_8\gamma)&=&a_1(T_{b3})^{[ij]}T(\bar{3})^k(T_8)_{[ij]k}+a_2(T_{b3})^{[ij]}T(3)^k(T_8)_{[ik]j},\label{Eq:HTb32T8}
\\
A(T_{b6} \to T_{8}\gamma)&=&a'_1(T_{b6})^{ij}T(\bar{3})^k(T_{8})_{[ik]j},\label{Eq:HTb62T8}\\
A(T_{b6} \to T_{10}\gamma)&=&a''_1(T_{b6})^{ij}T(\bar{3})^k(T_{10})_{ijk},\label{Eq:HTb62T10}
\end{eqnarray}
with  $T(\bar{3})=(0,1,1)$, which denotes the transition operators $(\bar{q}_2b)$  with $q_2=d,s$. The coefficients $a^{(','')}_i$, which contain information about QCD dynamics,  include the single quark emission contributions in Fig. \ref{Fig:q1tq2r} and  $T_{b3} \to T_{8}\psi_i$ and $T_{b6} \to T_{8,10}\psi_i$ ($\psi_i$ are the set of all $J=1,l=0 ~(c\bar{c})$) long distance contributions \cite{Dery:2020lbc,Deshpande:1994cn} (the similar in following $b^{(','')}_i,~ c^{(','',''')}_i,~ d_i $ in Eqs. (\ref{Eq:HTc32T8}-\ref{Eq:HT102T8ES})), nevertheless,  the long distance contributions in the
b-sector are small and under control \cite{Golowich:1994zr,Deshpande:1994cn}.
Noted that $T_{b3} \to T_{10}\gamma$ (and later $T_{c3} \to T_{10}\gamma)$ weak decays are not allowed by  the quark symmetry.
The SU(3) IRA amplitudes of $T_{b3}\to T_{8}\gamma $ and $T_{b6}\to T_{8,10}\gamma $ weak decays are given in Tab. \ref{Tab:HTb36tT810}. And the information of relevant CKM matrix elements  $V_{tb}V^*_{ts}$ ($V_{tb}V^*_{td}$ ) for $b\to s \gamma$ ($b\to d \gamma$) transition is not shown in Tab. \ref{Tab:HTb36tT810}.
\begin{table}[hb]
\renewcommand\arraystretch{1.15}
\tabcolsep 0.5in
\centering
\caption{The SU(3) IRA amplitudes of the  $T_{b3,b6}\to T_{8,10}\gamma $ weak decays by the $b\to s/d \gamma$ transitions, and $A_1\equiv a_1+a_2$. }\vspace{0.1cm}
{\footnotesize
\begin{tabular}{lc}  \hline\hline
Decay modes~~~~~~~~~~~~& $A(T_{b3,b6}\to T_{8,10}\gamma)$  \\\hline
{\color{blue}\bf $T_{b3}\to T_{8}\gamma $ via the $b\to s\gamma $ transition:}\\
$\Lambda^0_b\to \Lambda^0 \gamma $&$-2A_1/\sqrt{6}$\\
$\Lambda^0_b\to \Sigma^0 \gamma $& $0$\\
$\Xi^0_b\to \Xi^0 \gamma $& $-A_1$\\
$\Xi^-_b\to \Xi^- \gamma $& $A_1$\\
{\color{blue}\bf $T_{b3}\to T_{8}\gamma $ via the $b\to d\gamma $ transition:}\\
$\Lambda^0_b\to n \gamma $& $A_1$\\
$\Xi^0_b\to \Lambda^0 \gamma $&$-A_1/\sqrt{6}$ \\
$\Xi^0_b\to \Sigma^0 \gamma $&$-A_1/\sqrt{2}$ \\
$\Xi^-_b\to \Sigma^- \gamma $&$A_1$ \\
\hline
{\color{blue}\bf $T_{b6}\to T_{8}\gamma $ via the $b\to s\gamma $ transition:}\\
$\Sigma^+_b\to \Sigma^+\gamma $&$-a'_1$\\
$\Sigma^0_b\to \Lambda^0\gamma $&$0$\\
$\Sigma^0_b\to \Sigma^0\gamma $&$-a'_1$\\
$\Sigma^-_b\to \Sigma^-\gamma $&$a'_1$\\
$\Xi^{*0}_b\to \Xi^0\gamma $&$-a'_1/\sqrt{2}$\\
$\Xi^{*-}_b\to \Xi^-\gamma $&$a'_1/\sqrt{2}$\\
{\color{blue}\bf $T_{b6}\to T_{8}\gamma $ via the $b\to d\gamma $ transition:}\\
$\Sigma^+_b\to p\gamma $&$a'_1$\\
$\Sigma^0_b\to n\gamma $&$a'_1$\\
$\Xi^{*0}_b\to \Lambda^0\gamma $&$-\frac{\sqrt{3}}{2}a'_1$\\
$\Xi^{*0}_b\to \Sigma^0\gamma $&$a'_1/2$\\
$\Xi^{*-}_b\to \Sigma^-\gamma $&$-a'_1/\sqrt{2}$\\
$\Omega_b\to \Xi^-\gamma $&$-a'_1$\\
\hline
{\color{blue}\bf $T_{b6}\to T_{10}\gamma $ via the $b\to s\gamma $ transition:}\\
$\Sigma^+_b\to \Sigma^{*+}\gamma $&$a''_1$\\
$\Sigma^0_b\to \Sigma^{*0}\gamma $&$a'_1/2$\\
$\Sigma^-_b\to \Sigma^{*-}\gamma $&$a''_1$\\
$\Xi^{*0}_b\to \Xi^{*0}\gamma $&$a''_1/\sqrt{2}$\\
$\Xi^{*-}_b\to \Xi^{*-}\gamma $&$a''_1/\sqrt{2}$\\
$\Omega_b\to \Omega\gamma $&$a''_1$\\
{\color{blue}\bf $T_{b6}\to T_{10}\gamma $ via the $b\to d\gamma $ transition:}\\
$\Sigma^+_b\to \Delta^+\gamma $&$a''_1$\\
$\Sigma^0_b\to \Delta^0\gamma $&$a'_1/\sqrt{2}$\\
$\Sigma^-_b\to \Delta^-\gamma $&$\sqrt{3}a''_1$\\
$\Xi^{*0}_b\to \Sigma^{*0}\gamma $&$a''_1/2$\\
$\Xi^{*-}_b\to \Sigma^{*-}\gamma $&$a''_1/\sqrt{2}$\\
$\Omega_b\to \Xi^{*-}\gamma $&$a''_1$\\
\hline
\end{tabular}\label{Tab:HTb36tT810}}
\end{table}

\begin{table}[t]
\renewcommand\arraystretch{1.2}
\tabcolsep 0.2in
\centering
\caption{Branching ratios of the $T_{b3}\to T_{8}\gamma$ decays via the  $b\to s/d\gamma$ transitions. }\vspace{0.1cm}
{\footnotesize
\begin{tabular}{lccc}  \hline\hline
Observables~~~~~~~~~~~~& Experimental data \cite{PDG2020} &  Our SU(3) IRA predictions & Other predictions \\\hline
{\color{blue}\bf $b\to s:$}&\\
$\mathcal{B}(\Lambda^{0}_b\to \Lambda\gamma)(\times10^{-6})$&$7.1\pm1.7$&$7.1\pm3.4$ &$7.3\pm1.5$ \cite{Wang:2008sm} \\
$\mathcal{B}(\Lambda^{0}_b\to \Sigma^0\gamma)$&$\cdots$&$0$ & \\
$\mathcal{B}(\Xi^{-}_b\to \Xi^-\gamma)(\times10^{-5})$&$\cdots$&$1.23\pm0.64$   \\
$\mathcal{B}(\Xi^{0}_b\to \Xi^0\gamma)(\times10^{-5})$&$\cdots$&$1.16\pm0.60$   \\
{\color{blue}\bf $b\to d:$}&\\
$\mathcal{B}(\Lambda^{0}_b\to n\gamma)(\times10^{-7})$&$\cdots$&$5.03\pm2.67$&$^{3.69^{+3.76}_{-1.95}}_{3.7} $ \cite{Liu:2019rpm,Faustov:2020thr}\\
$\mathcal{B}(\Xi^{0}_b\to \Lambda^0\gamma)(\times10^{-8})$&$\cdots$&$9.17\pm5.10$   \\
$\mathcal{B}(\Xi^{0}_b\to \Sigma^0\gamma)(\times10^{-7})$&$\cdots$&$2.71\pm1.50$   \\
$\mathcal{B}(\Xi^{-}_b\to \Sigma^-\gamma)(\times10^{-7})$&$\cdots$&$5.74\pm3.21$   \\
\hline
\end{tabular}\label{Tab:Tb32T8}}
%\end{table}
\vspace{1cm}
%\begin{table}[ht]
\renewcommand\arraystretch{1.2}
\tabcolsep 0.5in
\centering
\caption{The SU(3) IRA amplitudes of the $T_{b3,b6}\to T_{c3,c6}\gamma $ weak decays by the $W$ exchange $b+u\to c+s/d+\gamma$. }\vspace{0.1cm}
{\footnotesize
\begin{tabular}{lc}  \hline\hline
Decay modes~~~~~~~~~~~~& $A(T_{b3,b6}\to T_{c3,c6}\gamma)$  \\\hline
{\color{blue}\bf $T_{b3}\to T_{c3}\gamma$:}\\
$\Xi^0_b\to \Xi^0_c \gamma $&$\widetilde{a}_1$\\
$\Lambda^0_b\to \Xi^0_c \gamma $&$-\widetilde{a}_1~s_c$\\\hline
{\color{blue}\bf $T_{b3}\to T_{c6}\gamma$:}\\
$\Lambda^0_b\to \Sigma^0_c \gamma $&$-\widetilde{a}'_1~$\\
$\Xi^0_b\to \Xi^{*0}_c \gamma $&$-\widetilde{a}'_1/\sqrt{2}$\\
$\Lambda^0_b\to \Xi^{*0}_c \gamma $&$-\widetilde{a}'_1~s_c/\sqrt{2}$\\
$\Xi^0_b\to \Omega_c \gamma $&$-\widetilde{a}'_1~s_c$\\\hline
{\color{blue}\bf $T_{b6}\to T_{c3}\gamma$:}\\
$\Sigma^+_b\to \Lambda^+_c \gamma $&$\widetilde{a}''_1$\\
$\Xi^{*0}_b\to \Xi^0_c \gamma $&$-\widetilde{a}''_1/\sqrt{2}$\\
$\Sigma^+_b\to \Xi^+_c \gamma $&$\widetilde{a}''_1~s_c$\\
$\Sigma^0_b\to \Xi^0_c \gamma $&$\widetilde{a}''_1~s_c/\sqrt{2}$\\\hline
{\color{blue}\bf $T_{b6}\to T_{c6}\gamma$:}\\
$\Sigma^+_b\to \Sigma^+_c \gamma $&$\widetilde{a}'''_1/\sqrt{2}$\\
$\Sigma^0_b\to \Sigma^0_c \gamma $&$\widetilde{a}'''_1/\sqrt{2}$\\
$\Xi^{*0}_b\to \Xi^{*0}_c \gamma $&$\widetilde{a}'''_1/2$\\
$\Sigma^+_b\to \Xi^{*+}_c \gamma $&$\widetilde{a}'''_1~s_c/\sqrt{2}$\\
$\Sigma^0_b\to\Xi^{*0}_c \gamma $&$\widetilde{a}'''_1~s_c/2$\\
$\Xi^{*0}_b\to \Omega_c \gamma $&$\widetilde{a}'''_1~s_c/\sqrt{2}$\\\hline
\end{tabular}\label{Tab:HTb36tTc36}}
\end{table}

Now, we discuss the  $T_{b3}\to T_{8}\gamma$ weak decays.
Decays $\Lambda^0_b\to \Lambda^0 \gamma $, $\Lambda^0_b\to \Sigma^0 \gamma $, $\Xi^0_b\to \Xi^0 \gamma $ and  $\Xi^-_b\to \Xi^- \gamma$ ($\Lambda^0_b\to n \gamma $, $\Xi^0_b\to \Lambda^0 \gamma $, $\Xi^0_b\to \Sigma^0 \gamma $ and $\Xi^-_b\to \Sigma^- \gamma$) proceed via the $b\to s\gamma$ ($b\to d\gamma$) flavor changing neutral current transition. From Tab. \ref{Tab:HTb36tT810}, one can see that the 7 decay amplitudes of  $T_{b3}\to T_{8}\gamma$  can be related by one parameter $A_1$.
Noted that  our $|A(T_{b3}\to T_{8}\gamma)|$ via the $b\to s/d\gamma$ transitions are consistent with ones of $T_{b3}\to T_{8}J/\psi$ in Ref. \cite{Fayyazuddin:2017sxq} and ones of the CKM-leading part results  $T_{b3}\to T_{8}J/\psi$ in Ref. \cite{Dery:2020lbc}.
Among these 7 decay modes, only $\mathcal{B}(\Lambda^{0}_b\to \Lambda\gamma)$  has been  measured at present, which is listed in the second column of Tab. \ref{Tab:Tb32T8}.  Using data of $\mathcal{B}(\Lambda^{0}_b\to \Lambda\gamma)$ and the expression of $\mathcal{B}(\mathcal{B}_1\to \mathcal{B}_2\gamma)$ in  Eq. (\ref{EQ:BrB1tB2r810}) to get $|A_1|$, and then  other 6 branching ratios are obtained, which are given in the third column of Tab. \ref{Tab:Tb32T8}.  Previous predictions for $\mathcal{B}(\Lambda^{0}_b\to \Lambda\gamma)$ in the light-cone sum rules and  for $\mathcal{B}(\Lambda^{0}_b\to n\gamma)$  in the relativistic quark model and in the Bethe-Salpeter equation approach are listed in the last column of Tab. \ref{Tab:Tb32T8}.  Our SU(3) IRA prediction of  $\mathcal{B}(\Lambda^{0}_b\to n\gamma)$  agrees with ones in  the relativistic quark model or in the Bethe-Salpeter equation approach \cite{Liu:2019rpm,Faustov:2020thr}. More experimental data about the $T_{b3}\to T_{8}\gamma$ decays in the further could test the SU(3) flavor symmetry approach.

As given in Tab. \ref{Tab:HTb36tT810},  the decay amplitudes of the $T_{b6}\to T_{8}\gamma$ and  $T_{b6}\to T_{10}\gamma$ weak decays can be parameterized by only one parameter $a'_1$ and $a''_1$, respectively.
 And  our  $|A(T_{b6}\to T_{8}\gamma)|$  via the $b\to s\gamma$ transition are consistent with ones of $T_{b6}\to T_{8}J/\psi$ in Ref. \cite{Fayyazuddin:2017sxq}.
 Unfortunately,  none  of the $T_{b6}\to T_{8}\gamma$ and  $T_{b6}\to T_{10}\gamma$ weak decays has been measured at present.  Any measurement of the $T_{b6}\to T_{8}\gamma$ ($T_{b6}\to T_{10}\gamma$) will give us chance to predict other 10 (11) decay modes.

In addition, $T_{b3,b6}$ baryons also can decay to $T_{c3,c6}$ by only  $W$ exchange $b+u\to c+s/d+\gamma$ \cite{Cheng:1994kp}.
The SU(3) IRA decay amplitudes for the $T_{b3,b6}\to T_{c3,c6}\gamma$ decays are
\begin{eqnarray}
A(T_{b3} \to T_{c3}\gamma)&=&\widetilde{a}_1V_{cb}V_{q_jq_l}(T_{b3})^{[ij]}(T_{c3})_{[il]},\label{Eq:HTb32Tc3}\\
A(T_{b3} \to T_{c6}\gamma)&=&\widetilde{a}'_1V_{cb}V_{q_jq_l}(T_{b3})^{[ij]}(T_{c6})_{il},\label{Eq:HTb32Tc6}\\
A(T_{b6} \to T_{c3}\gamma)&=&\widetilde{a}''_1V_{cb}V_{q_jq_l}(T_{b6})^{ij}(T_{c3})_{[il]},\label{Eq:HTb62Tc3}\\
A(T_{b6} \to T_{c6}\gamma)&=&\widetilde{a}'''_1V_{cb}V_{q_jq_l}(T_{b6})^{ij}(T_{c6})_{il}.\label{Eq:HTc62Tc6}
\end{eqnarray}
Since the CKM matrix element $V_{cb}$ occur for all processes, we will absorb it in the coefficients $\widetilde{a}_1,\widetilde{a}'_1,\widetilde{a}''_1$ as well as  $\widetilde{a}'''_1$, and only keep $V_{ud}\approx1$ and
$V_{us}\approx\lambda=s_c\equiv\mbox{sin}\theta_C\approx0.22453$ representing the Cabbibo angle $\theta_C$ \cite{PDG2020}.
The SU(3) IRA amplitudes of $T_{b3,b6}\to T_{c3,c6}\gamma$ weak decays are summarized in Tab. \ref{Tab:HTb36tTc36}, in which we keep $V_{ud}$ and $V_{us}$ information for  comparing conveniently, and  we may easily see the amplitude relations in this table.
Just none of these decays has been measured yet.

\subsection{$T_{c3,c6}$ weak radiative decays}
$T_{c3,c6}\to T_{8,10}\gamma$ via $c\to u\gamma$ transition are similar to the $T_{b3,b6}\to T_{8,10}\gamma$ via $b\to s/d\gamma$ transition.  Nevertheless, the short distance contributions from $c\to u \gamma$ could be negligible, and  the dominant contributions in charmed baryon decays mainly from the W-exchange contributions similar as Fig. \ref{fig:othersdr} (a-b) \cite{Singer:1995is}.
The SU(3) flavor structure of the relevant $b\to s,d$ Hamiltonian can been found, for instance,  in Refs. \cite{Zeppenfeld:1980ex,Savage:1989ub,Deshpande:1994ii}. The  SU(3) IRA amplitudes of the $T_{c3,c6}$ baryon weak decays are
\begin{eqnarray}
A(T_{c3} \to T_8\gamma)&=&b_1(T_{c3})^{[ij]}T'(\bar{3})^k(T_8)_{[ij]k}+b_2(T_{c3})^{[ij]}T'(3)^k(T_8)_{[ik]j}\nonumber\\
&&+\Big(\widetilde{b}_1H(\bar{6})^{lk}_j+\widetilde{b}_4H(15)^{lk}_j\Big)(T_{c3})^{[ij]}(T_8)_{[il]k}+\Big(\widetilde{b}_2H(\bar{6})^{lk}_j+\widetilde{b}_5H(15)^{lk}_j\Big)(T_{c3})^{[ij]}(T_8)_{[ik]l}\nonumber\\
&&+\Big(\widetilde{b}_3H(\bar{6})^{lk}_j+\widetilde{b}_6H(15)^{lk}_j\Big)(T_{c3})^{[ij]}(T_8)_{[lk]i},\label{Eq:HTc32T8}
\\
A(T_{c6} \to T_{8}\gamma)&=&b'_1(T_{c6})^{ij}T'(\bar{3})^k(T_{8})_{[ik]j}\nonumber\\
&&+\Big(\widetilde{b}'_1H(\bar{6})^{lk}_j+\widetilde{b}'_4H(15)^{lk}_j\Big)(T_{c6})^{ij}(T_8)_{[il]k}+\Big(\widetilde{b}'_2H(\bar{6})^{lk}_j+\widetilde{b}'_5H(15)^{lk}_j\Big)(T_{c6})^{ij}(T_8)_{[ik]l}\nonumber\\
&&+\Big(\widetilde{b}'_3H(\bar{6})^{lk}_j+\widetilde{b}'_6H(15)^{lk}_j\Big)(T_{c6})^{ij}(T_8)_{[lk]i},\label{Eq:HTc62T8}
\\ A(T_{c6} \to T_{10}\gamma)&=&b''_1(T_{c6})^{ij}T'(\bar{3})^k(T_{10})_{ijk}+\widetilde{b}''_1H(15)^{lk}_j(T_{c6})^{ij}(T_{10})_{ilk},
\label{Eq:HTc62T10}
\end{eqnarray}
with  $T'(\bar{3})=(1,0,0)$ which denotes the transition operators $(\bar{q}_2c)$  with $q_2=u$.
The $b^{(','')}_i$ terms denote the contributions from $c\to u \gamma$ shown in Fig.\ref{Fig:q1tq2r} and  the long distance contribution from the real $q\bar{q}$ intermediate
state $\rho$, $\omega$ and $\phi$.
The $\widetilde{b}^{(','')}_i$ terms denote the $W$ exchange contributions similar as Fig. \ref{fig:othersdr}. Noted that  $H(\bar{6})^{lk}_j$ ($H(15)^{lk}_j$) related to $(\bar{q}_lq^j)(\bar{q}_kc)$ operator  is antisymmetric (symmetric) in upper indices.
The non-vanish $H(\bar{6})^{lk}_j$  and $H(15)^{lk}_j$ for $c\to su\bar{d},du\bar{s},u\bar{d}d,u\bar{s}s$ transitions can be found in Ref. \cite{Wang:2017azm}.
Using $l,k$ antisymmetric in $H(\bar{6})^{lk}_j$ and $l,k$ symmetric in $H(15)^{lk}_j$,  we have
\begin{eqnarray}
\widetilde{b}^{(')}_2=-\widetilde{b}^{(')}_1, ~~~\widetilde{b}^{(')}_5=\widetilde{b}^{(')}_4, ~~~\widetilde{b}^{(')}_6=0.
\end{eqnarray}

Since none of the down-type quarks are heavy,  the Glashow-Iliuopoulos-Maiani (GIM) mechanism suppression is obvious in the charm sector.
The $b^{(','')}_i$ terms related to the short and long distance contributions of $c\to u \gamma$ transition are strongly suppressed by the GIM mechanism.
%CKM matrix elements $V_{cb}V_{ub}\approx A^2\lambda^5(\rho-i \eta)$.
As for the $W$ exchange transition,
there are three kinds of charm quark decays into light quarks
\begin{eqnarray}
d+c\to u+s+\gamma,~~~~~~  d+c\to u+d+\gamma~~(s+c\to u+s+\gamma),~~~~~~~s+c\to u+d+\gamma, \label{Eq:3kcdecays}
\end{eqnarray}
 which are related to
 $H(\bar{6},15)^{13}_2$, $H(\bar{6},15)^{12}_2~(H(\bar{6},15)^{13}_3)$, and  $H(\bar{6},15)^{12}_3$ are proportional to $V^*_{cs}V_{ud}$,  $V^*_{cd}V_{ud}$ ($V^*_{cs}V_{us}$), and  $V^*_{cd}V_{us}$, respectively. The relevant CKM matrix elements can be written by the Wolfenstein parameterization \cite{PDG2020}
\begin{eqnarray}
&&V^*_{cs}V_{ud}=(1-\lambda^2/2)^2\approx1,\nonumber\\
&&V^*_{cd}V_{ud}=-\lambda(1-\lambda^2/2)\approx-s_c,  ~~~~~~~V^*_{cs}V_{us}=\lambda(1-\lambda^2/2)\approx s_c,\nonumber\\
&&V^*_{cd}V_{us}=-\lambda^2 (1-\lambda^2/2)\approx-s_c^2.
\end{eqnarray}
So three kinds of charm quark decays in Eq. (\ref{Eq:3kcdecays}) are  called Cabibbo
allowed, singly Cabibbo suppressed, and doubly Cabibbo suppressed decays,  respectively.

The SU(3) IRA amplitudes of $T_{c3}\to T_{8}\gamma $ and $T_{c6}\to T_{8,10}\gamma$ weak decays are given in the second column of Tab. \ref{Tab:HTc36tT810}. For well understanding, we also show the relevant  CKM matrix information in Tab. \ref{Tab:HTc36tT810}, too.
In addition,
the contribution of $H(\bar{6})$ to the decay branching ratio can be about 5.5 times larger than
one of $H(15)$ due to Wilson Coefficient suppressed, for examples, see  \cite{Geng:2017esc,Geng:2018plk}.
If ignoring the GIM  strongly suppressed $c\to u \gamma$ transition contributions and the Wilson Coefficient suppressed $H(15)$ term contributions, the decay amplitudes of  $T_{c3}\to T_{8}\gamma $, $T_{c6}\to T_{8}\gamma $ and $T_{c6}\to T_{10}\gamma$
 are related by only one parameter  $\widetilde{B}\equiv\widetilde{b}_1-\widetilde{b}_3$, $\widetilde{B}'\equiv\widetilde{b}'_1-\widetilde{b}'_3$ and $\widetilde{b}''_1$, respectively.
The simplifications resulting are listed in the last column of Tab. \ref{Tab:HTc36tT810}.
Just all baryon weak radiative decays of $T_{c3,c6}$ baryons have not been measured yet.

\begin{table}[htb]
\renewcommand\arraystretch{1.}
\tabcolsep 0.15in
\centering
\caption{The SU(3) IRA amplitudes of the $T_{c3,c6}\to T_{8,10}\gamma $ weak decays,  $B_1\equiv b_1+b_2$, $\widetilde{B}_1\equiv\widetilde{b}_1-\widetilde{b}_3+\widetilde{b}_4$, $\widetilde{B}_2\equiv\widetilde{b}_1-\widetilde{b}_3-\widetilde{b}_4$, $\widetilde{B}'_1\equiv\widetilde{b}'_1-\widetilde{b}'_3+\widetilde{b}'_4$, $\widetilde{B}'_2\equiv\widetilde{b}'_1-\widetilde{b}'_3-\widetilde{b}'_4$ and $\widetilde{B}^{(')}\equiv\widetilde{b}^{(')}_1-\widetilde{b}^{(')}_3$. }\vspace{0.1cm}
{\footnotesize
\begin{tabular}{lcc}  \hline\hline
Decay modes~~~~~~~~~~~~& $\lambda_{q_1q_2}A(T_{c3,c6}\to T_{8,10}\gamma )$& approximative $\lambda_{q_1q_2}A(T_{c3,c6}\to T_{8,10}\gamma )$ \\\hline
{\color{blue}\bf  Cabibbo
allowed $T_{c3}\to T_{8}\gamma $:}\\
$\Lambda^+_c \to \Sigma^+\gamma$ & $-\widetilde{B}_1$& $-\widetilde{B}$\\
$\Xi^0_c \to \Xi^0\gamma$ & $-\widetilde{B}_2$& $-\widetilde{B}$\\
{\color{blue}\bf  singly Cabibbo suppressed $T_{c3}\to T_{8}\gamma $:}\\
$\Lambda^+_c\to p \gamma $& $\big[B_1-\big(\frac{5}{8}\widetilde{B}_1-\frac{1}{8}\widetilde{B}_2\big)\big]s_c$& $-\frac{1}{2}\widetilde{B}s_c$\\
$\Xi^+_c\to \Sigma^+ \gamma $&$\big[-B_1-\big(\frac{5}{8}\widetilde{B}_1-\frac{1}{8}\widetilde{B}_2\big)\big]s_c$& $-\frac{1}{2}\widetilde{B}s_c$\\
$\Xi^0_c\to \Lambda^0 \gamma $&$\big[B_1+3\big(\frac{1}{8}\widetilde{B}_1-\frac{5}{8}\widetilde{B}_2\big)\big]s_c/\sqrt{6}$ & $-\frac{3}{2}\widetilde{B}s_c/\sqrt{6}$ \\
$\Xi^0_c\to \Sigma^0 \gamma $&$\big[-B_1+\big(\frac{1}{8}\widetilde{B}_1-\frac{5}{8}\widetilde{B}_2\big)\big]s_c/\sqrt{2}$& $-\frac{1}{2}\widetilde{B}s_c/\sqrt{2}$ \\
{\color{blue}\bf doubly Cabibbo suppressed $T_{c3}\to T_{8}\gamma $:}\\
$\Xi^+_c\to p\gamma$&$\widetilde{B}_1s^2_c$&$\widetilde{B}s^2_c$\\
$\Xi^0_c\to n\gamma$&$\widetilde{B}_2s^2_c$&$\widetilde{B}s^2_c$\\
\hline
{\color{blue}\bf  Cabibbo
allowed $T_{c6}\to T_{8}\gamma $:}\\
$\Sigma^+_c\to \Sigma^+ \gamma $& $-\widetilde{B}'_1/\sqrt{2}$& $-\widetilde{B}'/\sqrt{2}$\\
$\Sigma^0_c\to \Lambda^0 \gamma $& $-\widetilde{B}'_2/\sqrt{6}$& $-\widetilde{B}'/\sqrt{6}$\\
$\Sigma^0_c\to \Sigma^0 \gamma $& $-\widetilde{B}'_1/\sqrt{2}$& $-\widetilde{B}'/\sqrt{2}$\\
$\Xi^{*0}_c\to \Xi^0 \gamma $& $-\widetilde{B}'_2/\sqrt{2}$& $-\widetilde{B}'/\sqrt{2}$\\
{\color{blue}\bf   singly Cabibbo suppressed $T_{c6}\to T_{8}\gamma $:}\\
$\Sigma^+_c\to p \gamma $& $\big[-b'_1-\big(\frac{5}{8}\widetilde{B}'_1-\frac{1}{8}\widetilde{B}'_2\big)\big]s_c/\sqrt{2}$ & $-\frac{1}{2}\widetilde{B}'s_c/\sqrt{2}$\\
$\Sigma^0_c\to n \gamma $& $\big[-2b'_1+\big(\frac{1}{8}\widetilde{B}'_1-\frac{5}{8}\widetilde{B}'_2\big)\big]s_c/\sqrt{2}$& $-\frac{1}{2}\widetilde{B}'s_c/\sqrt{2}$\\
$\Xi^{*+}_c\to \Sigma^+ \gamma $& $\big[b'_1-\big(\frac{5}{8}\widetilde{B}'_1-\frac{1}{8}\widetilde{B}'_2\big)\big]s_c/\sqrt{2}$& $-\frac{1}{2}\widetilde{B}'s_c/\sqrt{2}$\\
$\Xi^{*0}_c\to \Lambda^0 \gamma $&$\big[3b'_1-\big(\frac{13}{8}\widetilde{B}'_1-\frac{5}{8}\widetilde{B}'_2\big)\big]s_c/(2\sqrt{3})$
&$-\widetilde{B}'s_c/(2\sqrt{3})$\\
$\Xi^{*0}_c\to \Sigma^0 \gamma $&$\big[b'_1-\big(\frac{13}{8}\widetilde{B}'_1-\frac{5}{8}\widetilde{B}'_2\big)\big]s_c/2$
&$-\frac{1}{2}\widetilde{B}'s_c$\\
$\Omega^0_c\to \Xi^0 \gamma $&$\big[2b'_1+\big(\frac{1}{8}\widetilde{B}'_1-\frac{5}{8}\widetilde{B}'_2\big)\big]s_c$
&$-\frac{1}{2}\widetilde{B}'s_c$\\
{\color{blue}\bf   doubly Cabibbo suppressed $T_{c6}\to T_{8}\gamma $:}\\
$\Xi^{*+}_c\to p \gamma $&$\widetilde{B}'_1s^2_c/\sqrt{2}$&$\widetilde{B}'s^2_c/\sqrt{2}$\\
$\Xi^{*0}_c\to n \gamma $&$\widetilde{B}'_1s^2_c/\sqrt{2}$&$\widetilde{B}'s^2_c/\sqrt{2}$\\
$\Omega^0_c\to \Lambda^0 \gamma $&$-(\widetilde{B}'_1+\widetilde{B}'_2)s^2_c/\sqrt{6}$
&$-2\widetilde{B}'s^2_c/\sqrt{6}$\\
$\Omega^0_c\to \Sigma^0 \gamma $&$(\widetilde{B}'_1-\widetilde{B}'_2)s^2_c/\sqrt{2}$&0\\
\hline
{\color{blue}\bf  Cabibbo
allowed $T_{c6}\to T_{10}\gamma $:}\\
$\Sigma^+_c\to \Sigma^{*+} \gamma $& $\widetilde{b}''_1/\sqrt{2}$& $\widetilde{b}''_1/\sqrt{2}$\\
$\Sigma^0_c\to \Sigma^{*0} \gamma $& $\widetilde{b}''_1/\sqrt{2}$& $\widetilde{b}''_1/\sqrt{2}$\\
$\Xi^{*0}_c\to \Xi^{*0} \gamma $& $\widetilde{b}''_1/\sqrt{2}$& $\widetilde{b}''_1/\sqrt{2}$\\
{\color{blue}\bf   singly Cabibbo suppressed $T_{c6}\to T_{10}\gamma $:}\\
$\Sigma^{++}_{c}\to \Delta^{++} \gamma $& $\sqrt{3} b''_1s_c$&$0$\\
$\Sigma^+_c\to \Delta^+ \gamma $& $\big(b''_1-\frac{3}{4}\widetilde{b}''_1\big)s_c/\sqrt{2}$&$-\frac{3}{4}\widetilde{b}''_1s_c/\sqrt{2}$\\
$\Sigma^0_c\to \Delta^0 \gamma $&$\big(b''_1-\frac{3}{4}\widetilde{b}''_1\big)s_c$&$-\frac{3}{4}\widetilde{b}''_1s_c$ \\
$\Xi^{*+}_c\to \Sigma^{*+} \gamma $&$\big(b''_1+\frac{3}{4}\widetilde{b}''_1\big)s_c/\sqrt{2}$ &$\frac{3}{4}\widetilde{b}''_1s_c/\sqrt{2}$\\
$\Xi^{*0}_c\to \Sigma^{*0} \gamma $&$b''_1s_c/2$&$0$ \\
$\Omega_c\to \Xi^{*0} \gamma $&$\big(b''_1+\frac{3}{4}\widetilde{b}''_1\big)s_c$&$\frac{3}{4}\widetilde{b}''_1s_c$ \\
{\color{blue}\bf   doubly Cabibbo suppressed $T_{c6}\to T_{10}\gamma $:}\\
$\Sigma^+_c\to \Sigma^{*+} \gamma $& $\widetilde{b}''_1s^2_c/\sqrt{2}$& $\widetilde{b}''_1s^2_c/\sqrt{2}$\\
$\Sigma^0_c\to \Sigma^{*0} \gamma $& $\widetilde{b}''_1s^2_c/\sqrt{2}$& $\widetilde{b}''_1s^2_c/\sqrt{2}$\\
$\Xi^{*0}_c\to \Xi^{*0} \gamma $& $\widetilde{b}''_1s^2_c/\sqrt{2}$& $\widetilde{b}''_1s^2_c/\sqrt{2}$\\
\hline
\end{tabular}\label{Tab:HTc36tT810}}
\end{table}

%\newpage

\subsection{$T_{8,10}$ weak radiative decays}

The SU(3) flavor structure of the relevant $s\to d$ Hamiltonian can be found in Ref. \cite{Wang:2019alu}.
The decay amplitudes  of the $T_8$ and $T_{10}$  weak radiative  decays can be parameterized as
\begin{eqnarray}
A(T_{8} \to T'_8\gamma)&=&c_1(T_{8})^{[ij]n}T''(\bar{3})^k(T'_8)_{[ij]k}+c_2(T_{8})^{[ij]n}T'(3)^k(T_8)_{[ik]j}\nonumber\\
&+&c_3(T_{8})^{[in]j}T''(\bar{3})^k(T'_8)_{[ij]k}+c_4(T_{8})^{[in]j}T''(\bar{3})^k(T'_8)_{[ik]j}+c_5(T_{8})^{[in]j}T''(\bar{3})^k(T'_8)_{[jk]i} \nonumber\\
&+&\widetilde{c}_1(T_{8})^{[ij]n}(T'_8)_{[il]k}H(4)^{lk}_j+\widetilde{c}_2(T_{8})^{[in]j}(T'_8)_{[il]k}H(4)^{lk}_j+\widetilde{c}_3(T_{8})^{[jn]i}(T'_8)_{[il]k}H(4)^{lk}_j,\label{Eq:HT82T8p}\\
A(T_{8} \to T'_{10}\gamma)&=&c'_1(T_{8})^{[in]j}T''(\bar{3})^k(T_{10})_{ijk}\nonumber\\
&+&\widetilde{c}'_1(T_{8})^{[in]j}(T_{10})_{ilk}H(4)^{lk}_j+\widetilde{c}'_2(T_{8})^{[jn]i}(T_{10})_{ilk}H(4)^{lk}_j+\widetilde{c}'_3(T_{8})^{[ij]n}(T_{10})_{ilk}H(4)^{lk}_j,\label{Eq:HT82T10}\\
A(T_{10} \to T_{8}\gamma)&=&c''_1(T_{10})^{ijn}T''(\bar{3})^k(T_{8})_{[ik]j}+\widetilde{c}''_1(T_{10})^{ijn}(T'_8)_{[il]k}H(4)^{lk}_j,\label{Eq:HT102T8}
\\A(T_{10} \to T'_{10})&=&c'''_1(T_{10})^{ijn}T''(\bar{3})^k(T'_{10})_{ijk}+\widetilde{c}'''_1(T_{10})^{ijn}(T_{10})_{ilk}H(4)^{lk}_j,\label{Eq:HT102T10p}
\end{eqnarray}
where $n\equiv 3$ for $s$ quark, $T''(\bar{3})=(0,1,0)$ related to the transition operator $(\bar{d}s)$, and $H(4)^{lk}_j$ related to $(\bar{q}_lq^j)(\bar{q}_ks)$ operator  is symmetric in upper indices \cite{Wang:2019alu}. In Eqs. (\ref{Eq:HT82T8p}-\ref{Eq:HT102T10p}), the $c^{(','',''')}_i$ terms denote the contributions from $s\to d \gamma$ shown in Fig.\ref{Fig:q1tq2r} and  the long distance contribution from $\psi_i,~\rho$ and $\omega$ \cite{Singer:1995is}  (for $T_{8,10}$ weak radiative decays, the long distance contributions may be significantly larger than the short distance ones),   the $\widetilde{c}^{(','',''')}_i$ terms denote the $W$ exchange contributions shown in   Fig. \ref{fig:othersdr} (a-b), and the internal radiation contributions in Fig. \ref{fig:othersdr} (c) are neglected in this work.

\newpage
The SU(3) IRA amplitudes of the $T_{8,10}\to T^{'}_{8,10}\gamma $ weak decays are summarized  in Tab. \ref{Tab:HT810tT810W}, in which
the information of the same CKM matrix elements  $V_{us}V^*_{ud}$ is not shown.
\begin{table}[b]
\renewcommand\arraystretch{1.2}
\tabcolsep 0.5in
\centering
\caption{The SU(3) IRA amplitudes of the $T_{8,10}\to T^{'}_{8,10}\gamma $ weak decays, $C_1\equiv c_1+c_2+c_3-c_5$ and $C_2\equiv c_4+c_5$,   $\widetilde{C}_A\equiv \widetilde{c}_1-\widetilde{c}_3$,  $\widetilde{C}_B\equiv \widetilde{c}_2+\widetilde{c}_3$, $\widetilde{C}'_A\equiv \widetilde{c}'_1+\widetilde{c}'_2$ and  $\widetilde{C}'_B\equiv \widetilde{c}'_2+\widetilde{c}'_3$.}\vspace{0.1cm}
{\footnotesize
\begin{tabular}{lc}  \hline\hline
Decay modes~~~~~~~~~~~~& $A(T_{8,10}\to T^{'}_{8,10}\gamma)$  \\\hline
{\color{blue}\bf $T_{8}\to T'_{8}\gamma $ weak decays:}&\\
$\Xi^-\to \Sigma^- \gamma $&$C_1$\\
$\Xi^0\to \Lambda^0  \gamma $&$(C_1+2C_2)/\sqrt{6}$\\
$\Xi^0\to \Sigma^0  \gamma $&$(C_1+2\widetilde{C}_A)/\sqrt{2}$\\
$\Lambda^0\to n \gamma $&$-(2C_1+C_2+2\widetilde{C}_A+\widetilde{C}_B)/\sqrt{6}$ \\
$\Sigma^+\to p \gamma $&$-(C_2+\widetilde{C}_B)$ \\
$\Sigma^0\to n \gamma $&$-(C_2-\widetilde{C}_B)/\sqrt{2}$\\\hline
{\color{blue}\bf $T_{8}\to T_{10}\gamma $ weak decays:}&\\
$\Xi^0\to \Sigma^{*0}  \gamma $&$(-c'_1+\widetilde{C}'_B)/\sqrt{2}$\\
$\Xi^-\to \Sigma^{*-} \gamma $&$c'_1$\\
$\Sigma^-\to \Delta^-  \gamma $&$2\sqrt{3}c'_1$\\
$\Sigma^+\to \Delta^+ \gamma $&$-(2c'_1+\widetilde{C}'_A)$ \\
$\Lambda^0\to \Delta^0 \gamma $&$(\widetilde{C}'_A+\widetilde{C}'_B)/\sqrt{6}$ \\
$\Sigma^0\to \Delta^0 \gamma $&$-(2c'_1+\widetilde{C}'_A)/\sqrt{2}$\\\hline
{\color{blue}\bf $T_{10}\to T_{8}\gamma $ weak decays:}&\\
$\Omega^-\to \Xi^-  \gamma $&$-c''_1$\\
$\Xi^{*-}\to \Sigma^- \gamma $&$-c''_1/\sqrt{3}$\\
$\Xi^{*0}\to \Lambda^0  \gamma $&$-c''_1/\sqrt{2}$\\
$\Xi^{*0}\to \Sigma^0  \gamma $&$(c''_1+2\widetilde{c}''_1)/\sqrt{6}$\\
$\Sigma^{*0}\to n \gamma $&$(c''_1-\widetilde{c}''_1)/\sqrt{6}$ \\
$\Sigma^{*+}\to p \gamma $&$(c''_1+\widetilde{c}''_1)/\sqrt{3}$\\\hline
{\color{blue}\bf $T_{10}\to T'_{10}\gamma $ weak decays:}&\\
$\Omega^-\to \Xi^{*-}  \gamma $&$\sqrt{3}c'''_1$\\
$\Xi^{*-}\to \Sigma^{*-} \gamma $&$c'''_1$\\
$\Xi^{*0}\to \Sigma^{*0}  \gamma $&$(c'''_1+\widetilde{c}'''_1)/\sqrt{2}$\\
$\Sigma^{*+}\to \Delta^+  \gamma $&$c'''_1+\widetilde{c}'''_1$\\
$\Sigma^{*0}\to \Delta^0 \gamma $&$(c'''_1+\widetilde{c}'''_1)/\sqrt{2}$ \\
$\Sigma^{*-}\to \Delta^-\gamma $&$\sqrt{3}c'''_1$\\\hline
\end{tabular}\label{Tab:HT810tT810W}}
\end{table}
From Tab. \ref{Tab:HT810tT810W},
one can see that the amplitudes of
$\Xi^-\to \Sigma^- \gamma $, $\Xi^-\to \Sigma^{*-} \gamma $, $\Sigma^-\to \Delta^-  \gamma $, $\Omega^-\to \Xi^-  \gamma $,
$\Xi^{*-}\to \Sigma^- \gamma $, $\Omega^-\to \Xi^{*-}  \gamma $,
$\Xi^{*-}\to \Sigma^{*-} \gamma $, $\Sigma^{*-}\to \Delta^-\gamma $ only contain coefficients $c^{(','',''')}_i$, which means that the $W$ exchange transitions don't contribute to these decays  since the initial baryons don't contain $u$ quark. Otherwise, the $W$ exchange contributions are canceled in $\Xi^0\to \Lambda^0  \gamma $ and $\Xi^{*0}\to \Lambda^0  \gamma$ decays. So above decays could be used to explore the short distance and long distance contributions.
Other decay amplitudes contained both $c^{(','',''')}_i$ and $\widetilde{c}^{(','',''')}_i$ could proceed from the short  distance contributions, long distance contributions and   W-exchange contributions.
\begin{table}[b]
\renewcommand\arraystretch{1.2}
\tabcolsep 0.25in
\centering
\caption{Branching ratios of the $T_{8,10}\to T_{8}\gamma$ weak decays via $s\to d\gamma$ transition. }\vspace{0.1cm}
{\footnotesize
\begin{tabular}{lccc}  \hline\hline
Observables~~~~~~~~~~~~& Experimental data \cite{PDG2020} &  Our  predictions in $S_A$& Our predictions in $S_B$ \\\hline
{\color{blue}\bf $T_{8}\to T'_{8}\gamma$:}&\\
$\mathcal{B}(\Xi^{-}\to \Sigma^-\gamma)(\times10^{-4})$&$1.27\pm0.23$&  $1.67\pm0.06$&$1.27\pm0.46$ \\
$\mathcal{B}(\Xi^{0}\to \Lambda\gamma)(\times10^{-3})$&$1.17\pm0.07$&$3.05\pm0.29$ &$1.17\pm0.14$  \\
$\mathcal{B}(\Xi^{0}\to \Sigma^0\gamma)(\times10^{-3})$&$3.33\pm0.10$&  $0.14\pm0.01$ &$3.33\pm0.20$  \\
$\mathcal{B}(\Lambda^0\to n\gamma)(\times10^{-3})$&$1.75\pm0.15$& $1.48\pm0.03$ &$1.75\pm0.30$  \\
$\mathcal{B}(\Sigma^{+}\to p\gamma)(\times10^{-3})$&$1.23\pm0.05$&  $1.27\pm0.06$ & $1.23\pm0.10$ \\
$\mathcal{B}(\Sigma^0\to n\gamma)(\times10^{-13})$&$\cdots$& $6.12\pm1.35$&$3.146\pm3.143$ \\
\hline
\end{tabular}\label{Tab:T8T8rW}}
\end{table}

For the weak $T_{8}\to T'_{8}\gamma$ decays,
all decay modes expect for $\Sigma^0\to n\gamma$ have been measured and paid a lot of attentions, experimental data are listed in the second column of Tab. \ref{Tab:T8T8rW},  and there are longstanding theoretical difficulties to explain the experimental data of  the weak $T_{8}\to T'_{8}\gamma$ decays.
In the $T_{8}\to T'_{8}\gamma$ weak decays, they may decay via the $s\to d\gamma$
single quark emission, corresponding  long distance effects and the $W$-exchange transition.  Since the quarks are   antisymmetric in  both the initial states $T_8$  and the final states $T'_8$, there are more independent parameters than ones in $T_{b3,c3}\to T_{8}\gamma$ weak decays.
The relevant SU(3) flavor parameters could be complex, and we set $C_1$ is  real and add relative phases $\delta_{11}$, $\delta_{1A}$, $\delta_{1B}$  for $C_2$, $\widetilde{C}_A$  and $\widetilde{C}_B$, respectively. And then 7 independent parameters given by
\begin{eqnarray}
C_1, ~C_2e^{i\delta_{12}},~\widetilde{C}_Ae^{i\delta_{1A}},~ \widetilde{C}_Be^{i\delta_{1B}}.
\end{eqnarray}
Two cases for the $T_{8}\to T'_{8}\gamma$ weak decays will be considered in our analysis. In case $S_A$,  we will only consider the $s\to d\gamma$
single quark emission and long distance effects, $i.e.$, set $\widetilde{C}_A=\widetilde{C}_B=0$. In case $S_B$,  we will consider all effects.

In case $S_A$, there are 3 independent parameters $C_1$ and $C_2e^{i\delta_{12}}$. Firstly, we use three data of $\mathcal{B}(\Xi^{-}\to \Sigma^-\gamma)$, $\mathcal{B}(\Lambda^0\to n\gamma)$ and $\mathcal{B}(\Sigma^{+}\to p\gamma)$ to constrain the parameters as well as  obtain that $C_1=32.27\pm6.23$, $C_2=48.96\pm9.74$ and $\delta_{12}=(0.08\pm21.58)^\circ$, and noted that $C_2$ is slightly larger than $C_1$.
Then, we use the obtained $C_1$, $C_2$ and $\delta_{12}$ to predict other three branching ratios, and the results are given in the third column of Tab. \ref{Tab:T8T8rW}.  One can see that the predictions   of $\mathcal{B}(\Xi^{0}\to \Lambda\gamma)$  and $\mathcal{B}(\Xi^{0}\to \Sigma^0\gamma)$ in $S_A$ are not inconsistent  with their data.   In case $S_A$, since  $\frac{A(\Xi^0\to \Sigma^0\gamma)}{A(\Xi^-\to \Sigma^-\gamma)}=\frac{1}{\sqrt{2}}$,
 $m_{\Xi^0}\approx m_{\Xi^-}$, $m_{\Sigma^0}\approx m_{\Sigma^-}$ and $\frac{\tau_{\Xi^0}}{\tau_{\Xi^-}}\approx 1.8$, one have that $\frac{\mathcal{B}(\Xi^0\to \Sigma^0\gamma)}{\mathcal{B}(\Xi^-\to \Sigma^-\gamma)}\approx0.9$, which are far away from the experimental one $\frac{\mathcal{B}^{Exp.}(\Xi^0\to \Sigma^0\gamma)}{\mathcal{B}^{Exp}(\Xi^-\to \Sigma^-\gamma)}\approx26.2$. In addition, the SU(3) IRA prediction of $\mathcal{B}(\Xi^{0}\to \Lambda\gamma)$ is about 2.6 times larger than its data.  So it's necessary to considering the $W$ exchange contributions.

In case $S_B$, we use all five data of the branching ratios to constrain seven parameters. We obtain that $C_1=28.04\pm8.77$, $C_2=41.32\pm21.95$,  $\widetilde{C}_A=77.82\pm24.29$, $\widetilde{C}_B=45.82\pm45.48$, $\delta_{12}=(-3.05\pm176.86)^\circ$, $\delta_{1A}=(4.30\pm171.46)^\circ$  and $\delta_{1B}=(-0.03\pm179.36)^\circ$. One can see that, after satisfying all present data within $2\sigma$, three phases $\delta_{12},\delta_{1A},\delta_{1B}$ are almost unlimited, and the $C_1$, $C_2$ and $\widetilde{C}_A$ terms give the same magnitude contributions. $\widetilde{C}_B$ lies in $[0.33,91.30]$, and its contribution might be similar to (or smaller than) ones of $C_1$, $C_2$ and $\widetilde{C}_A$.

In both  $S_A$ and $S_B$ cases,  the branching ratio of $\Sigma^0\to n\gamma$ is too small to see in the experiments, and this weak decay completely overwhelmed by the  simpler electromagnetic decay $\Sigma^0\to \Lambda^0\gamma$.

For the baryon decuplet radiative weak decays, only $\Omega$ baryon has a sufficiently long lifetime,   is's accessible to experimental study. Using the experimental upper limit  $\mathcal{B}(\Omega^-\to \Xi\gamma)<4.60\times10^{-4}$, we obtain that   $\mathcal{B}(\Xi^{*-}\to \Sigma^-\gamma)<1.82\times10^{-16}$ and $\mathcal{B}(\Xi^{*0}\to \Lambda^0\gamma)<3.53\times10^{-16}$.  The upper limits of the two branching ratios are very tiny since $\Xi^{*0,-}$ have  very short lifetime,  so the $\Xi^{*-}\to \Sigma^-\gamma$ and $\Xi^{*0}\to \Lambda^0\gamma$ decays are difficult to be measured in the experiments.

\subsection{$T_{10}\to T_8\gamma$  electromagnetic radiative decays}

In addition, the  baryon decuplet $T_{10}$ also can decay only through electromagnetic interactions by the $q_n\to q_{n'}\gamma(n=n')$ transition at the quark level.  The  SU(3) IRA amplitudes of the $T_{10} \to T_8\gamma$ electromagnetic decays can be parameterized as
\begin{eqnarray}
A^E(T_{10} \to T_8\gamma)&=&d_1(T_{10})^{ijn}(T_8)_{[in']j}.\label{Eq:HT102T8ES}
\end{eqnarray}
Three cases are considered in calculating the electromagnetic decay amplitudes. $S_1$: assuming all three quarks in $T_{10}$ baryon can emit photon,  $S_2$: assuming $d$ and $s$ quarks in $T_{10}$ baryon can emit photon, and $S_3$: assuming the heaviest quark in $T_{10}$ baryon can emit photon.  The  SU(3) IRA amplitudes of $T_{10} \to T_8\gamma$ electromagnetic decays in three cases  are listed in Tab. \ref{Tab:HT810tT810E}.

\begin{table}[b]
\renewcommand\arraystretch{1.2}
\tabcolsep 0.16in
\centering
\caption{The SU(3) IRA amplitudes of $T_{10}\to T_{8}\gamma $ electromagnetic decays. }\vspace{0.1cm}
{\footnotesize
\begin{tabular}{lccc}  \hline\hline
Decay modes~~~~~& $A^{E}(T_{10}\to T_{8}\gamma )$ in $S_1$  & $A^{E}(T_{10}\to T_{8}\gamma )$ in $S_2$& $A^{E}(T_{10}\to T_{8}\gamma )$ in $S_3$ \\\hline
$\Xi^{*-}\to \Xi^- \gamma $&$0$&$0$ &$-2d_1/\sqrt{3}$  \\\
$\Xi^{*0}\to \Xi^0 \gamma $&$0$&$-2d_1/\sqrt{3}$ &$-2d_1/\sqrt{3}$ \\
$\Delta^{+}\to p \gamma $&$0$&$2d_1/\sqrt{3}$ &$2d_1/\sqrt{3}$ \\
$\Delta^{0}\to n \gamma $&$0$&$2d_1/\sqrt{3}$ &$2d_1/\sqrt{3}$ \\
$\Sigma^{*-}\to \Sigma^- \gamma $&$0$&$0$ &$2d_1/\sqrt{3}$ \\
$\Sigma^{*0}\to \Lambda^0 \gamma $&$0$&$-d_1$ &$0$\\
$\Sigma^{*0}\to \Sigma^0 \gamma $&$0$&$-d_1/\sqrt{3}$ &$-2d_1/\sqrt{3}$  \\
$\Sigma^{*+}\to \Sigma^+ \gamma $&$0$&$-2d_1/\sqrt{3}$ &$-2d_1/\sqrt{3}$ \\
\hline
\end{tabular}\label{Tab:HT810tT810E}}
\end{table}
For $T_{10}\to T_{8}\gamma $ electromagnetic decays, $\mathcal{B}(\Sigma^{*+}\to \Sigma^+\gamma)$ and $\mathcal{B}(\Sigma^{*0}\to \Lambda^0\gamma)$ have been measured, $\mathcal{B}(\Xi^{*}\to \Xi\gamma)$ and $\mathcal{B}(\Sigma^{*-}\to \Sigma^-\gamma)$ have been upper limited, and the relevant experimental data are listed in the second column of Tab. \ref{Tab:BrT10T8rES}.   Comparing the amplitudes in three cases with the data, the $S_1$ and $S_3$ cases are eliminated, and we will use the IRA amplitudes in $S_2$ case  in the  following analysis.

Using Eq. (\ref{EQ:BrB1tB2rES}), the branching ratios will be obtained by $\mathcal{M}(T_{10}\to T_8\gamma)=A^E(T_{10}\to T_8\gamma)$, and the results are listed in the last column of Tab. \ref{Tab:BrT10T8rES}. The SU(3) IRA predictions for  $\mathcal{B}(\Sigma^{*+}\to \Sigma^+\gamma)$ and $\mathcal{B}(\Sigma^{*0}\to \Lambda^0\gamma)$  are quite  consistent with present data. Noted that the estimated result $B(\Delta\to N\gamma)=(5.5-6.5)\times10^{-3}$ from PDG \cite{PDG2020}, which is covered by our prediction.
 The branching ratio predictions are at the order of $10^{-3}$, and they might be measured at the BESIII or LHC experiments in near future.  So these $T_{10}\to T_{8}\gamma $ electromagnetic decays could be used to test the SU(3) flavor symmetry.

\begin{table}[t]
\renewcommand\arraystretch{1.2}
\tabcolsep 0.3in
\centering
\caption{Branching ratios of $T_{10}\to T_{8}\gamma$ electromagnetic decays. }\vspace{0.1cm}
{\footnotesize
\begin{tabular}{lcc}  \hline\hline
Observables~~~~~~~~~~~~& Experimental data \cite{PDG2020} &   Our predictions \\\hline
%
%{\color{blue}\bf $T_{8}\to T'_{8}\gamma$ ES decay:}&\\
% $\mathcal{B}(\Sigma^{0}\to \Lambda\gamma)$&$100\%$& \\\hline
%
%{\color{blue}\bf $T_{10}\to T_{8}\gamma$ ES decays:}&\\
%
$\mathcal{B}(\Xi^{*}\to \Xi\gamma)(\times10^{-3})$&$<37$&$2.64\pm1.06$   \\
$\mathcal{B}(\Sigma^{*+}\to \Sigma^+\gamma)(\times10^{-3})$&$7.0\pm1.7$&$6.48\pm2.80$   \\
$\mathcal{B}(\Sigma^{*0}\to \Lambda^0\gamma)(\times10^{-2})$&$1.25^{+0.13}_{-0.12}$&$1.26\pm0.25$    \\
$\mathcal{B}(\Sigma^{*0}\to \Sigma^{0}\gamma)(\times10^{-3})$&$\cdots$&$1.57\pm0.31$   \\
$\mathcal{B}(\Sigma^{*-}\to \Sigma^-\gamma)(\times10^{-4})$&$<2.4$&$0$   \\
$\mathcal{B}(\Delta^0\to n\gamma)(\times10^{-3})$&$\cdots$&$8.25\pm3.64$ \\
$\mathcal{B}(\Delta^+\to p\gamma)(\times10^{-3})$&$\cdots$&$8.36\pm3.68$  \\\hline
\end{tabular}\label{Tab:BrT10T8rES}}
\end{table}

\section{Conclusions}
Baryon radiative decays give us a chance to study the interplay of the electromagnetic, weak and strong interactions.
Some baryon radiative decay  modes have been measured and some others could be studied at BESIII, LHCb and Belle-II experiments.
In this work, we have analyzed baryon radiative decays of the
octet $T_{8}$,  decuplet $T_{10}$,  single charmed anti-triplet $T_{c3}$, single charmed  sextet $T_{c6}$,  single  bottomed anti-triplet $T_{b3}$ and single bottomed sextet $T_{b6}$
by using the irreducible representation approach to test the SU(3) flavor symmetry.
Our main results are given in order:
\begin{itemize}
\item{\bf $T_{b3,b6}$ weak radiative decays}: \\
Each kind of the decay amplitudes can be  related by only one parameter in the $T_{b3}\to T_{8}\gamma$, $T_{b6}\to T_{8}\gamma$ and $T_{b6}\to T_{10}\gamma$  weak decays via $b\to s/d \gamma$ as well as the $T_{b3}\to T_{c3}\gamma$, $T_{b3}\to T_{c6}\gamma$, $T_{b6}\to T_{c3}\gamma$ and $T_{b6}\to T_{c6}\gamma$ via the $W$ exchange transitions. Using the only measured $\mathcal{B}(\Lambda^{0}_b\to \Lambda\gamma)$, we have predicted other six decay branching ratios of $T_{b3}\to T_{8}\gamma$ weak decays, and they might be measured by the experiments in near future.
Unfortunately,  none  of $T_{b6}\to T_{8,10}\gamma$ and   $T_{b3,b6}\to T_{c3,c6}\gamma$ weak decays has been measured at present.  Any measurement of $T_{b6}\to T_{8}\gamma,T_{10}\gamma$ will give us chance to predict many other decay branching ratios.

\item{\bf $T_{c3,c6}$ weak radiative decays}: \\
$T_{c3,c6}$ weak radiative decays are quite different from $T_{b3,b6}$ weak radiative decays, they may receive the contributions of both the $c\to u\gamma$  and the $W$ exchange transitions. After ignoring the GIM  strongly suppressed $c\to u \gamma$ transition contributions and the Wilson Coefficient suppressed $H(15)$ term contributions, the decay amplitudes of  $T_{c3}\to T_{8}\gamma $, $T_{c6}\to T_{8}\gamma $ and $T_{c6}\to T_{10}\gamma$
 are also related by only one parameter  $\widetilde{B}$, $\widetilde{B}'$ and $\widetilde{b}''_1$, respectively. Just none of $T_{c3,c6}$ weak radiative decays has been measured at present.

\item{\bf $T_{8,10}$ weak radiative decays}:\\
As for $T_{8,10}$ weak radiative decays, some decays only receive the short distance and long distance  $s\to d\gamma$ transition contributions, other  decays  could receive  both the $s\to d\gamma$ transition  and the W-exchange transition contributions.
For the  $T_{8}\to T'_{8}\gamma$ weak decays,
all decay modes expect for $\Sigma^0\to n\gamma$ have been measured, and we have found that only considering the short  and long distance  $s\to d\gamma$ transition contributions can not explain all current data by SU(3) IRA. Present all  data could be explained by considering both the  $s\to d\gamma$ transition contributions and the $W$ exchange contributions. $\mathcal{B}(\Sigma^0\to n\gamma)$ has been predicted, just this branching ratio is very tiny. For the $T_{10}\to T_{8}\gamma$ weak decays,
we have used the upper limit of $\mathcal{B}(\Omega^-\to \Xi\gamma)$  to obtain the upper limit predictions of $\mathcal{B}(\Xi^{*-}\to \Sigma^-\gamma)$ and $\mathcal{B}(\Xi^{*0}\to \Lambda^0\gamma)$, just  both upper limit predictions are tiny.

\item{\bf $T_{10}\to T_8\gamma$ electromagnetic radiative decays}:\\
All decay amplitudes of $T_{10}\to T_8\gamma$ electromagnetic radiative decays could be  related by only one parameter  $d_1$, the SU(3) IRA predictions for  $\mathcal{B}(\Sigma^{*+}\to \Sigma^+\gamma)$ and $\mathcal{B}(\Sigma^{*0}\to \Lambda^0\gamma)$  are quite consistent with present data.  Other branching ratio predictions are at the order of $10^{-3}$, and they might be measured by the experiments in near future.

\end{itemize}

Flavor SU(3) symmetry could   provide us very useful
information about the decays.
According to our predictions, some branching ratios are accessible to the experiments at BESIII, LHCb and Belle-II. Our results in this work can be used to test SU(3) flavor symmetry approach in the radiative baryon decays by the future experiments.

\section*{ACKNOWLEDGEMENTS}
The work was supported by the National Natural Science Foundation of China (Contract No. 11675137) and the Key Scientific Research Projects of Colleges and Universities in Henan Province (Contract No. 18A140029).

\section*{References}


\begin{thebibliography}{99}

%\cite{Cerri:2018ypt}
\bibitem{Cerri:2018ypt}
  A.~Cerri {\it et al.},
  %``Report from Working Group 4 : Opportunities in Flavour Physics at the HL-LHC and HE-LHC,''
  CERN Yellow Rep.\ Monogr.\  {\bf 7}, 867 (2019)
  % doi:10.23731/CYRM-2019-007.867
  [arXiv:1812.07638 [hep-ph]].
  %%CITATION = doi:10.23731/CYRM-2019-007.867;%%
  %104 citations counted in INSPIRE as of 11 Aug 2020


\bibitem{Aaij:2017ddf}
  R.~Aaij {\it et al.} (LHCb Collaboration),
  %``Evidence for the rare decay $\Sigma^+ \to p \mu^+ \mu^-$,''
  Phys.\ Rev.\ Lett.\  {\bf 120}, no. 22, 221803 (2018)
  % doi:10.1103/PhysRevLett.120.221803
  [arXiv:1712.08606 [hep-ex]].
  %%CITATION = % doi:10.1103/PhysRevLett.120.221803;%%
  %9 citations counted in INSPIRE as of 13 May 2019

%\cite{Junior:2018odx}
\bibitem{Junior:2018odx}
  A.~A.~Alves Junior {\it et al.},
  %``Prospects for Measurements with Strange Hadrons at LHCb,''
  JHEP {\bf 1905}, 048 (2019),
%  doi:10.1007/JHEP05(2019)048
  [arXiv:1808.03477 [hep-ex]].
 %%CITATION = doi:10.1007/JHEP05(2019)048;%%
 %7 citations counted in INSPIRE as of 18 Jun 2019


\bibitem{Aaij:2019hhx}
  R.~Aaij {\it et al.} (LHCb Collaboration),
  %``First Observation of the Radiative Decay  $\Lambda_{b}^{0} \to \Lambda \gamma$,''
  Phys.\ Rev.\ Lett.\  {\bf 123}, no. 3, 031801 (2019)
 % doi:10.1103/PhysRevLett.123.031801
  [arXiv:1904.06697 [hep-ex]].
  %%CITATION = doi:10.1103/PhysRevLett.123.031801;%%
  %9 citations counted in INSPIRE as of 08 Aug 2020


\bibitem{PDG2020}
P.A. Zyla et al. (Particle Data Group), to be published in Prog. Theor. Exp. Phys. 2020, 083C01 (2020).



\bibitem{Lach:1995we}
  J.~Lach and P.~Zenczykowski,
  %``Hyperon radiative decays,''
  Int.\ J.\ Mod.\ Phys.\ A {\bf 10}, 3817 (1995).
 % doi:10.1142/S0217751X95001807
  %%CITATION = doi:10.1142/S0217751X95001807;%%
  %57 citations counted in INSPIRE as of 18 Jul 2020


\bibitem{Donoghue:1985rk}
  J.~F.~Donoghue, E.~Golowich and B.~R.~Holstein,
  %``Low-Energy Weak Interactions of Quarks,''
  Phys.\ Rept.\  {\bf 131}, 319 (1986).
  % doi:10.1016/0370-1573(86)90151-1
  %%CITATION = doi:10.1016/0370-1573(86)90151-1;%%
  %182 citations counted in INSPIRE as of 08 Aug 2020

\bibitem{Ablikim:2018zay}
  M.~Ablikim {\it et al.} (BESIII Collaboration),
  %``Polarization and Entanglement in Baryon-Antibaryon Pair Production in Electron-Positron Annihilation,''
  Nature Phys.\  {\bf 15}, 631 (2019)
  %doi:10.1038/s41567-019-0494-8
  [arXiv:1808.08917 [hep-ex]].
  %%CITATION = doi:10.1038/s41567-019-0494-8;%%
  %51 citations counted in INSPIRE as of 11 Aug 2020


\bibitem{Li:2016tlt}
  H.~B.~Li,
  %``Prospects for rare and forbidden hyperon decays at BESIII,''
  Front.\ Phys.\ (Beijing) {\bf 12}, no. 5, 121301 (2017)
  % doi:10.1007/s11467-017-0691-9
  [arXiv:1612.01775 [hep-ex]].
  %%CITATION = % doi:10.1007/s11467-017-0691-9;%%
  %10 citations counted in INSPIRE as of 16 Apr 2019

%\cite{Bigi:2017eni}
\bibitem{Bigi:2017eni}
  I.~I.~Bigi, X.~W.~Kang and H.~B.~Li,
  %``CP Asymmetries in Strange Baryon Decays,''
  Chin.\ Phys.\ C {\bf 42}, no. 1, 013101 (2018)
 % doi:10.1088/1674-1137/42/1/013101
  [arXiv:1704.04708 [hep-ph]].
  %%CITATION = doi:10.1088/1674-1137/42/1/013101;%%
  %6 citations counted in INSPIRE as of 18 Jun 2019

\bibitem{Asner:2008nq}
  D.~M.~Asner {\it et al.},
  %``Physics at BES-III,''
  Int.\ J.\ Mod.\ Phys.\ A {\bf 24}, S1 (2009)
  [arXiv:0809.1869 [hep-ex]].
  %%CITATION = ARXIV:0809.1869;%%
  %271 citations counted in INSPIRE as of 13 May 2019


\bibitem{Dery:2020lbc}
  A.~Dery, M.~Ghosh, Y.~Grossman and S.~Schacht,
  %``SU(3)$_{F}$ analysis for beauty baryon decays,''
  JHEP {\bf 2003}, 165 (2020)
  % doi:10.1007/JHEP03(2020)165
  [arXiv:2001.05397 [hep-ph]].
  %%CITATION = doi:10.1007/JHEP03(2020)165;%%
  %7 citations counted in INSPIRE as of 19 Jul 2020

\bibitem{He:1998rq}
  X.~G.~He,
  %``SU(3) analysis of annihilation contributions and CP violating relations in B ---> P P decays,''
  Eur.\ Phys.\ J.\ C {\bf 9}, 443 (1999)
  % doi:10.1007/s100529900064
  [hep-ph/9810397].
  %%CITATION = % doi:10.1007/s100529900064;%%
  %66 citations counted in INSPIRE as of 17 Apr 2019




\bibitem{He:2000ys}
  X.~G.~He, Y.~K.~Hsiao, J.~Q.~Shi, Y.~L.~Wu and Y.~F.~Zhou,
  %``The CP violating phase $\gamma$ from global fit of rare charmless hadronic $B$ decays,''
  Phys.\ Rev.\ D {\bf 64}, 034002 (2001)
  % doi:10.1103/PhysRevD.64.034002
  [hep-ph/0011337].
  %%CITATION = % doi:10.1103/PhysRevD.64.034002;%%
  %62 citations counted in INSPIRE as of 22 Mar 2019


\bibitem{Fu:2003fy}
  H.~K.~Fu, X.~G.~He and Y.~K.~Hsiao,
  %``B ---> eta(eta-prime) K(pi) in the standard model with flavor symmetry,''
  Phys.\ Rev.\ D {\bf 69}, 074002 (2004)
  % doi:10.1103/PhysRevD.69.074002
  [hep-ph/0304242].
  %%CITATION = % doi:10.1103/PhysRevD.69.074002;%%
  %44 citations counted in INSPIRE as of 16 Apr 2019

\bibitem{Hsiao:2015iiu}
  Y.~K.~Hsiao, C.~F.~Chang and X.~G.~He,
  %``A global $SU(3)/U(3)$ flavor symmetry analysis for $B\to PP$ with $\eta-\eta'$ Mixing,''
  Phys.\ Rev.\ D {\bf 93}, no. 11, 114002 (2016)
  % doi:10.1103/PhysRevD.93.114002
  [arXiv:1512.09223 [hep-ph]].
  %%CITATION = % doi:10.1103/PhysRevD.93.114002;%%
  %24 citations counted in INSPIRE as of 16 Apr 2019

\bibitem{He:2015fwa}
  X.~G.~He and G.~N.~Li,
  %``Predictive $CP$ violating relations for charmless two-body decays of beauty baryons $\Xi^{-,\;0}_b$ and $\Lambda_b^0$ with flavor $SU(3)$ symmetry,''
  Phys.\ Lett.\ B {\bf 750}, 82 (2015)
  % doi:10.1016/j.physletb.2015.08.048
  [arXiv:1501.00646 [hep-ph]].
  %%CITATION = % doi:10.1016/j.physletb.2015.08.048;%%
  %20 citations counted in INSPIRE as of 16 Apr 2019

\bibitem{Gronau:1994rj}
  M.~Gronau, O.~F.~Hernandez, D.~London and J.~L.~Rosner,
  %``Decays of B mesons to two light pseudoscalars,''
  Phys.\ Rev.\ D {\bf 50}, 4529 (1994)
  % doi:10.1103/PhysRevD.50.4529
  [hep-ph/9404283].
  %%CITATION = % doi:10.1103/PhysRevD.50.4529;%%
  %404 citations counted in INSPIRE as of 17 Apr 2019

\bibitem{Gronau:1995hm}
  M.~Gronau, O.~F.~Hernandez, D.~London and J.~L.~Rosner,
  %``Broken SU(3) symmetry in two-body B decays,''
  Phys.\ Rev.\ D {\bf 52}, 6356 (1995)
  % doi:10.1103/PhysRevD.52.6356
  [hep-ph/9504326].
  %%CITATION = % doi:10.1103/PhysRevD.52.6356;%%
  %214 citations counted in INSPIRE as of 17 Apr 2019

\bibitem{Zhou:2016jkv}
  S.~H.~Zhou, Q.~A.~Zhang, W.~R.~Lyu and C.~D.~L��,
  %``Analysis of Charmless Two-body B decays in Factorization Assisted Topological Amplitude Approach,''
  Eur.\ Phys.\ J.\ C {\bf 77}, no. 2, 125 (2017)
  % doi:10.1140/epjc/s10052-017-4685-0
  [arXiv:1608.02819 [hep-ph]].
  %%CITATION = % doi:10.1140/epjc/s10052-017-4685-0;%%
  %16 citations counted in INSPIRE as of 17 Apr 2019


\bibitem{Cheng:2014rfa}
  H.~Y.~Cheng, C.~W.~Chiang and A.~L.~Kuo,
  %``Updating B��PP,VP decays in the framework of flavor symmetry,''
  Phys.\ Rev.\ D {\bf 91}, no. 1, 014011 (2015)
  % doi:10.1103/PhysRevD.91.014011
  [arXiv:1409.5026 [hep-ph]].
  %%CITATION = % doi:10.1103/PhysRevD.91.014011;%%
  %34 citations counted in INSPIRE as of 17 Apr 2019



\bibitem{He:2015fsa}
  M.~He, X.~G.~He and G.~N.~Li,
  %``CP-Violating Polarization Asymmetry in Charmless Two-Body Decays of Beauty Baryons,''
  Phys.\ Rev.\ D {\bf 92}, no. 3, 036010 (2015)
  % doi:10.1103/PhysRevD.92.036010
  [arXiv:1507.07990 [hep-ph]].
  %%CITATION = % doi:10.1103/PhysRevD.92.036010;%%
  %12 citations counted in INSPIRE as of 16 Apr 2019

 \bibitem{Deshpande:1994ii}
  N.~G.~Deshpande and X.~G.~He,
  %``CP asymmetry relations between anti-b0 ---> pi pi and anti-b0 ---> pi K rates,''
  Phys.\ Rev.\ Lett.\  {\bf 75}, 1703 (1995)
  % doi:10.1103/PhysRevLett.75.1703
  [hep-ph/9412393].
  %%CITATION = % doi:10.1103/PhysRevLett.75.1703;%%
  %81 citations counted in INSPIRE as of 16 Apr 2019

  \bibitem{Shivashankara:2015cta}
  S.~Shivashankara, W.~Wu and A.~Datta,
  %``$\Lambda_b \to \Lambda_c \tau \bar{\nu}_{\tau}$ Decay in the Standard Model and with New Physics,''
  Phys.\ Rev.\ D {\bf 91},  115003 (2015)
  % doi:10.1103/PhysRevD.91.115003
  [arXiv:1502.07230 [hep-ph]].
  %CITATION = % doi:10.1103/PhysRevD.91.115003;%%
  %12 citations counted in INSPIRE as of 22 May 2017


\bibitem{Singer:1995is}
  P.~Singer,
  %``Weak radiative decays of hyperons and of charm and beauty baryons,''
  Nucl.\ Phys.\ Proc.\ Suppl.\  {\bf 50} (1996) 202
  % doi:10.1016/0920-5632(96)00392-1
  [hep-ph/9512308].
  %%CITATION = doi:10.1016/0920-5632(96)00392-1;%%
  %6 citations counted in INSPIRE as of 26 Jul 2020




  \bibitem{Grossman:2012ry}
  Y.~Grossman and D.~J.~Robinson,
  %``SU(3) Sum Rules for Charm Decay,''
  JHEP {\bf 1304}, 067 (2013)
  % doi:10.1007/JHEP04(2013)067
  [arXiv:1211.3361 [hep-ph]].
  %%CITATION = % doi:10.1007/JHEP04(2013)067;%%
  %43 citations counted in INSPIRE as of 22 Mar 2019


  \bibitem{Pirtskhalava:2011va}
  D.~Pirtskhalava and P.~Uttayarat,
  %``CP Violation and Flavor SU(3) Breaking in D-meson Decays,''
  Phys.\ Lett.\ B {\bf 712}, 81 (2012)
  % doi:10.1016/j.physletb.2012.04.039
  [arXiv:1112.5451 [hep-ph]].
  %%CITATION = % doi:10.1016/j.physletb.2012.04.039;%%
  %93 citations counted in INSPIRE as of 16 Apr 2019


\bibitem{Cheng:2012xb}
  H.~Y.~Cheng and C.~W.~Chiang,
  %``SU(3) symmetry breaking and CP violation in D -> PP decays,''
  Phys.\ Rev.\ D {\bf 86}, 014014 (2012)
  % doi:10.1103/PhysRevD.86.014014
  [arXiv:1205.0580 [hep-ph]].
  %%CITATION = % doi:10.1103/PhysRevD.86.014014;%%
  %54 citations counted in INSPIRE as of 16 Apr 2019

\bibitem{Savage:1989qr}
  M.~J.~Savage and R.~P.~Springer,
  %``SU(3) Predictions for Charmed Baryon Decays,''
  Phys.\ Rev.\ D {\bf 42}, 1527 (1990).
  % doi:10.1103/PhysRevD.42.1527
  %%CITATION = % doi:10.1103/PhysRevD.42.1527;%%
  %43 citations counted in INSPIRE as of 16 Apr 2019


\bibitem{Savage:1991wu}
  M.~J.~Savage,
  %``SU(3) violations in the nonleptonic decay of charmed hadrons,''
  Phys.\ Lett.\ B {\bf 257}, 414 (1991).
  % doi:10.1016/0370-2693(91)91917-K
  %%CITATION = % doi:10.1016/0370-2693(91)91917-K;%%
  %48 citations counted in INSPIRE as of 16 Apr 2019

\bibitem{Altarelli:1975ye}
  G.~Altarelli, N.~Cabibbo and L.~Maiani,
  %``Weak Nonleptonic Decays of Charmed Hadrons,''
  Phys.\ Lett.\  {\bf 57B}, 277 (1975).
  % doi:10.1016/0370-2693(75)90075-1
  %%CITATION = % doi:10.1016/0370-2693(75)90075-1;%%
  %82 citations counted in INSPIRE as of 16 Apr 2019

\bibitem{Lu:2016ogy}
  C.~D.~L\"{u}, W.~Wang and F.~S.~Yu,
  %``Test flavor SU(3) symmetry in exclusive $\Lambda_c$ decays,''
  Phys.\ Rev.\ D {\bf 93}, no. 5, 056008 (2016)
  % doi:10.1103/PhysRevD.93.056008
  [arXiv:1601.04241 [hep-ph]].
  %%CITATION = % doi:10.1103/PhysRevD.93.056008;%%
  %42 citations counted in INSPIRE as of 16 Apr 2019


\bibitem{Geng:2017esc}
  C.~Q.~Geng, Y.~K.~Hsiao, Y.~H.~Lin and L.~L.~Liu,
  %``Non-leptonic two-body weak decays of $\Lambda_c(2286)$,''
  Phys.\ Lett.\ B {\bf 776}, 265 (2018)
  % doi:10.1016/j.physletb.2017.11.062
  [arXiv:1708.02460 [hep-ph]].
  %%CITATION = % doi:10.1016/j.physletb.2017.11.062;%%
  %16 citations counted in INSPIRE as of 16 Apr 2019

\bibitem{Geng:2018plk}
  C.~Q.~Geng, Y.~K.~Hsiao, C.~W.~Liu and T.~H.~Tsai,
  %``Antitriplet charmed baryon decays with SU(3) flavor symmetry,''
  Phys.\ Rev.\ D {\bf 97}, no. 7, 073006 (2018)
  % doi:10.1103/PhysRevD.97.073006
  [arXiv:1801.03276 [hep-ph]].
  %%CITATION = % doi:10.1103/PhysRevD.97.073006;%%
  %15 citations counted in INSPIRE as of 16 Apr 2019

\bibitem{Geng:2017mxn}
  C.~Q.~Geng, Y.~K.~Hsiao, C.~W.~Liu and T.~H.~Tsai,
  %``Charmed Baryon Weak Decays with SU(3) Flavor Symmetry,''
  JHEP {\bf 1711}, 147 (2017)
  % doi:10.1007/JHEP11(2017)147
  [arXiv:1709.00808 [hep-ph]].
  %%CITATION = % doi:10.1007/JHEP11(2017)147;%%
  %18 citations counted in INSPIRE as of 16 Apr 2019


\bibitem{Geng:2019bfz}
  C.~Q.~Geng, C.~W.~Liu, T.~H.~Tsai and S.~W.~Yeh,
  %``Semileptonic decays of anti-triplet charmed baryons,''
  Phys.\ Lett.\ B {\bf 792}, 214 (2019)
  % doi:10.1016/j.physletb.2019.03.056
  [arXiv:1901.05610 [hep-ph]].
  %%CITATION = doi:10.1016/j.physletb.2019.03.056;%%
  %11 citations counted in INSPIRE as of 11 Aug 2020


\bibitem{Wang:2017azm}
  W.~Wang, Z.~P.~Xing and J.~Xu,
  %``Weak Decays of Doubly Heavy Baryons: SU(3) Analysis,''
  Eur.\ Phys.\ J.\ C {\bf 77}, no. 11, 800 (2017)
  % doi:10.1140/epjc/s10052-017-5363-y
  [arXiv:1707.06570 [hep-ph]].
  %%CITATION = % doi:10.1140/epjc/s10052-017-5363-y;%%
  %49 citations counted in INSPIRE as of 16 Apr 2019


\bibitem{Wang:2019dls}
  D.~Wang,
  %``Sum rules for $CP$ asymmetries of charmed baryon decays in the $SU(3)_F$ limit,''
  Eur.\ Phys.\ J.\ C {\bf 79}, no. 5, 429 (2019)
  % doi:10.1140/epjc/s10052-019-6925-y
  [arXiv:1901.01776 [hep-ph]].
  %%CITATION = doi:10.1140/epjc/s10052-019-6925-y;%%
  %10 citations counted in INSPIRE as of 11 Aug 2020

\bibitem{Wang:2017gxe}
  D.~Wang, P.~F.~Guo, W.~H.~Long and F.~S.~Yu,
  %``K asymmetries and CP violation in charmed baryon decays into neutral kaons,''
  JHEP {\bf 1803}, 066 (2018)
  % doi:10.1007/JHEP03(2018)066
  [arXiv:1709.09873 [hep-ph]].
  %%CITATION = % doi:10.1007/JHEP03(2018)066;%%
  %13 citations counted in INSPIRE as of 16 Apr 2019


\bibitem{Muller:2015lua}
  S.~M\"{u}ller, U.~Nierste and S.~Schacht,
  %``Topological amplitudes in $D$ decays to two pseudoscalars: A global analysis with linear $SU(3)_F$ breaking,''
  Phys.\ Rev.\ D {\bf 92}, no. 1, 014004 (2015)
  % doi:10.1103/PhysRevD.92.014004
  [arXiv:1503.06759 [hep-ph]].
  %%CITATION = % doi:10.1103/PhysRevD.92.014004;%%
  %38 citations counted in INSPIRE as of 17 Apr 2019


\bibitem{Xu:2020jfr}
  Y.~G.~Xu, X.~D.~Cheng, J.~L.~Zhang and R.~M.~Wang,
  %``Studying Two-body Nonleptonic Weak Decays of Hyperons with Topological Diagram Approach,''
  J.\ Phys.\ G {\bf 47}, no. 8, 085005 (2020)
  %doi:10.1088/1361-6471/ab97c7
  [arXiv:2001.06907 [hep-ph]].
  %%CITATION = doi:10.1088/1361-6471/ab97c7;%%

\bibitem{Wang:2019alu}
  R.~M.~Wang, M.~Z.~Yang, H.~B.~Li and X.~D.~Cheng,
  %``Testing SU(3) Flavor Symmetry in Semileptonic and Two-body Nonleptonic Decays of Hyperons,''
  Phys.\ Rev.\ D {\bf 100}, no. 7, 076008 (2019)
  % doi:10.1103/PhysRevD.100.076008
  [arXiv:1906.08413 [hep-ph]].
  %%CITATION = doi:10.1103/PhysRevD.100.076008;%%
  %4 citations counted in INSPIRE as of 19 Jul 2020

\bibitem{Chang:2014iba}
  H.~M.~Chang, M. G. Alonso and J.~M. Camalich,
  %``Nonstandard Semileptonic Hyperon Decays,''
  Phys.\ Rev.\ Lett.\  {\bf 114}, no. 16, 161802 (2015)
  % doi:10.1103/PhysRevLett.114.161802
  [arXiv:1412.8484 [hep-ph]].
  %%CITATION = doi:10.1103/PhysRevLett.114.161802;%%
  %19 citations counted in INSPIRE as of 11 Aug 2020

\bibitem{Zenczykowski:2005cs}
  P.~Zenczykowski,
  %``Joint description of weak radiative and nonleptonic hyperon decays in broken SU(3),''
  Phys.\ Rev.\ D {\bf 73}, 076005 (2006)
  % doi:10.1103/PhysRevD.73.076005
  [hep-ph/0512122].
  %%CITATION = doi:10.1103/PhysRevD.73.076005;%%
  %10 citations counted in INSPIRE as of 08 Aug 2020

\bibitem{Zenczykowski:2006se}
  P.~Zenczykowski,
  %``Radiative and nonleptonic hyperon decays in broken SU(3),''
  Nucl.\ Phys.\ Proc.\ Suppl.\  {\bf 167}, 54 (2007)
  % doi:10.1016/j.nuclphysbps.2006.12.043
  [hep-ph/0610191].
  %%CITATION = doi:10.1016/j.nuclphysbps.2006.12.043;%%
  %3 citations counted in INSPIRE as of 08 Aug 2020




























\bibitem{Bos:1996ig}
  J.~W.~Bos, D.~Chang, S.~C.~Lee, Y.~C.~Lin and H.~H.~Shih,
  %``Hyperon weak radiative decays in chiral perturbation theory,''
  Phys.\ Rev.\ D {\bf 54}, 3321 (1996)
  % doi:10.1103/PhysRevD.54.3321
  [hep-ph/9601299].
  %%CITATION = doi:10.1103/PhysRevD.54.3321;%%
  %9 citations counted in INSPIRE as of 08 Aug 2020


%\cite{He:2006ud}
\bibitem{He:2006ud}
  X.~G.~He, T.~Li, X.~Q.~Li and Y.~M.~Wang,
  %``PQCD calculation for Lambda(b) ---> Lambda gamma in the standard model,''
  Phys.\ Rev.\ D {\bf 74}, 034026 (2006)
  % doi:10.1103/PhysRevD.74.034026
  [hep-ph/0606025].
  %%CITATION = doi:10.1103/PhysRevD.74.034026;%%
  %42 citations counted in INSPIRE as of 08 Aug 2020


\bibitem{Singer:1996xh}
  P.~Singer and D.~X.~Zhang,
  %``Weak radiative decays of beauty baryons,''
  Phys.\ Lett.\ B {\bf 383}, 351 (1996)
  % doi:10.1016/0370-2693(96)00747-2
  [hep-ph/9606343].
  %%CITATION = doi:10.1016/0370-2693(96)00747-2;%%
  %7 citations counted in INSPIRE as of 18 Jul 2020


\bibitem{Liu:2019rpm}
  L.~L.~Liu, C.~Wang, X.~W.~Kang and X.~H.~Guo,
  %``FCNC transitions of $\Lambda_b$ to neutron in Bethe-Salpeter equation approach,''
  Eur.\ Phys.\ J.\ C {\bf 80}, no. 3, 193 (2020)
  % doi:10.1140/epjc/s10052-020-7667-6
  [arXiv:1912.12622 [hep-ph]].
  %%CITATION = doi:10.1140/epjc/s10052-020-7667-6;%%


\bibitem{Faustov:2017ous}
  R.~N.~Faustov and V.~O.~Galkin,
  %``Rare $\Lambda_b\to n l^+l^-$ decays in the relativistic quark-diquark picture,''
  Mod.\ Phys.\ Lett.\ A {\bf 32}, 1750125 (2017)
  % doi:10.1142/S0217732317501255
  [arXiv:1706.01379 [hep-ph]].
  %%CITATION = doi:10.1142/S0217732317501255;%%
  %4 citations counted in INSPIRE as of 11 Jul 2020


\bibitem{Aliev:2004ju}
  T.~M.~Aliev and A.~Ozpineci,
  %``Radiative decays of decuplet to octet baryons in light cone QCD,''
  Nucl.\ Phys.\ B {\bf 732}, 291 (2006)
  % doi:10.1016/j.nuclphysb.2005.07.038
  [hep-ph/0406331].
  %%CITATION = doi:10.1016/j.nuclphysb.2005.07.038;%%
  %29 citations counted in INSPIRE as of 15 Jul 2020


\bibitem{Colangelo:2007jy}
  P.~Colangelo, F.~De Fazio, R.~Ferrandes and T.~N.~Pham,
  %``FCNC $B_s$ and $Lambda_b$ transitions: Standard model versus a single universal extra dimension scenario,''
  Phys.\ Rev.\ D {\bf 77}, 055019 (2008)
  % doi:10.1103/PhysRevD.77.055019
  [arXiv:0709.2817 [hep-ph]].
  %%CITATION = doi:10.1103/PhysRevD.77.055019;%%
  %35 citations counted in INSPIRE as of 08 Aug 2020


\bibitem{Cheng:1994kp}
  H.~Y.~Cheng, C.~Y.~Cheung, G.~L.~Lin, Y.~C.~Lin, T.~M.~Yan and H.~L.~Yu,
  %``Effective Lagrangian approach to weak radiative decays of heavy hadrons,''
  Phys.\ Rev.\ D {\bf 51}, 1199 (1995)
  % doi:10.1103/PhysRevD.51.1199
  [hep-ph/9407303].
  %%CITATION = doi:10.1103/PhysRevD.51.1199;%%
  %86 citations counted in INSPIRE as of 08 Aug 2020


%\cite{Ramalho:2020tnn}
\bibitem{Ramalho:2020tnn}
  G.~Ramalho,
  %``A covariant model for the decuplet to octet Dalitz decays,''
  arXiv:2002.07280 [hep-ph].
  %%CITATION = ARXIV:2002.07280;%%
  %1 citations counted in INSPIRE as of 07 Aug 2020



\bibitem{Junker:2019vvy}
  O.~Junker, S.~Leupold, E.~Perotti and T.~Vitos,
  %``Electromagnetic form factors of the transition from the spin-3/2 $\Sigma$ to the $\Lambda$ hyperon,''
  Phys.\ Rev.\ C {\bf 101}, no. 1, 015206 (2020)
  % doi:10.1103/PhysRevC.101.015206
  [arXiv:1910.07396 [hep-ph]].
  %%CITATION = doi:10.1103/PhysRevC.101.015206;%%
  %1 citations counted in INSPIRE as of 05 Aug 2020

\bibitem{Buchalla:1995vs}
  G.~Buchalla, A.~J.~Buras and M.~E.~Lautenbacher,
  %``Weak decays beyond leading logarithms,''
  Rev.\ Mod.\ Phys.\  {\bf 68}, 1125 (1996)
  % doi:10.1103/RevModPhys.68.1125
  [hep-ph/9512380].
  %%CITATION = doi:10.1103/RevModPhys.68.1125;%%
  %2560 citations counted in INSPIRE as of 10 Jul 2020


\bibitem{Faustov:2017wbh}
  R.~N.~Faustov and V.~O.~Galkin,
  %``Rare $\Lambda_b\to\Lambda l^+l^-$ and $\Lambda_b\to\Lambda\gamma$ decays in the relativistic quark model,''
  Phys.\ Rev.\ D {\bf 96}, no. 5, 053006 (2017)
  % doi:10.1103/PhysRevD.96.053006
  [arXiv:1705.07741 [hep-ph]].
  %%CITATION = doi:10.1103/PhysRevD.96.053006;%%
  %13 citations counted in INSPIRE as of 11 Jul 2020


\bibitem{Gutsche:2013pp}
  T.~Gutsche, M.~A.~Ivanov, J.~G.~Korner, V.~E.~Lyubovitskij and P.~Santorelli,
  %``Rare baryon decays $\Lambda_b \to \Lambda {l^{+}l^{-}} (l=e, \mu, \tau)$ and $\Lambda_b \to \Lambda\gamma$ : differential and total rates, lepton- and hadron-side forward-backward asymmetries,''
  Phys.\ Rev.\ D {\bf 87}, 074031 (2013)
  % doi:10.1103/PhysRevD.87.074031
  [arXiv:1301.3737 [hep-ph]].
  %%CITATION = doi:10.1103/PhysRevD.87.074031;%%
  %77 citations counted in INSPIRE as of 11 Jul 2020



\bibitem{Mannel:2011xg}
  T.~Mannel and Y.~M.~Wang,
  %``Heavy-to-light baryonic form factors at large recoil,''
  JHEP {\bf 1112}, 067 (2011)
  % doi:10.1007/JHEP12(2011)067
  [arXiv:1111.1849 [hep-ph]].
  %%CITATION = doi:10.1007/JHEP12(2011)067;%%
  %53 citations counted in INSPIRE as of 26 Jul 2020



\bibitem{Verma:1988gf}
  R.~C.~Verma and A.~Sharma,
  %``A Reanalysis of Weak Radiative Decays of Hyperons,''
  Phys.\ Rev.\ D {\bf 38}, 1443 (1988).
  % doi:10.1103/PhysRevD.38.1443
  %%CITATION = doi:10.1103/PhysRevD.38.1443;%%
  %34 citations counted in INSPIRE as of 18 Jul 2020


\bibitem{Azimov:1996uf}
  Y.~I.~Azimov,
  %``Hara's theorem and W exchange in hyperon weak radiative decays,''
  Z.\ Phys.\ A {\bf 359}, 75 (1997)
  % doi:10.1007/s002180050369
  [hep-ph/9611315].
  %%CITATION = doi:10.1007/s002180050369;%%
  %7 citations counted in INSPIRE as of 18 Jul 2020


\bibitem{Dubovik:2007qg}
  E.~N.~Dubovik, V.~S.~Zamiralov and S.~N.~Lepshokov,
  %``Weak radiative hyperon decays in quark model,''
  AIP Conf.\ Proc.\  {\bf 964}, no. 1, 71 (2007)
  % doi:10.1063/1.2823883
  [hep-ph/0701141].
  %%CITATION = doi:10.1063/1.2823883;%%


\bibitem{Zeppenfeld:1980ex}
  D.~Zeppenfeld,
  %``SU(3) Relations for B Meson Decays,''
  Z.\ Phys.\ C {\bf 8}, 77 (1981).
  % doi:10.1007/BF01429835
  %%CITATION = doi:10.1007/BF01429835;%%
  %193 citations counted in INSPIRE as of 20 Jul 2020

\bibitem{Savage:1989ub}
  M.~J.~Savage and M.~B.~Wise,
  %``SU(3) Predictions for Nonleptonic B Meson Decays,''
  Phys.\ Rev.\ D {\bf 39}, 3346 (1989)
  Erratum: [Phys.\ Rev.\ D {\bf 40}, 3127 (1989)].
  % doi:10.1103/PhysRevD.39.3346, 10.1103/PhysRevD.40.3127
  %%CITATION = doi:10.1103/PhysRevD.39.3346, 10.1103/PhysRevD.40.3127;%%
  %168 citations counted in INSPIRE as of 20 Jul 2020


\bibitem{Deshpande:1994cn}
  N.~G.~Deshpande, X.~G.~He and J.~Trampetic,
  %``Long distance contributions to penguin processes b ---> s gamma and b ---> d gamma,''
  Phys.\ Lett.\ B {\bf 367}, 362 (1996)
  % doi:10.1016/0370-2693(95)01364-4
  [hep-ph/9412222].
  %%CITATION = doi:10.1016/0370-2693(95)01364-4;%%
  %106 citations counted in INSPIRE as of 18 Jul 2020

\bibitem{Golowich:1994zr}
  E.~Golowich and S.~Pakvasa,
  %``Uncertainties from long range effects in B ---> K* gamma,''
  Phys.\ Rev.\ D {\bf 51}, 1215 (1995)
  % doi:10.1103/PhysRevD.51.1215
  [hep-ph/9408370].
  %%CITATION = doi:10.1103/PhysRevD.51.1215;%%
  %86 citations counted in INSPIRE as of 18 Jul 2020


\bibitem{Fayyazuddin:2017sxq}
  Fayyazuddin and M.~J.~Aslam,
  %``Hadronic weak decay $\mathcal{B}_{b}(\frac{1}{2}^+) \to \mathcal{B}(\frac{1}{2}^{+},\; \frac{3}{2}^{+}) +V$,''
  Phys.\ Rev.\ D {\bf 95}, no. 11, 113002 (2017)
  % doi:10.1103/PhysRevD.95.113002
  [arXiv:1705.05106 [hep-ph]].
  %%CITATION = doi:10.1103/PhysRevD.95.113002;%%
  %5 citations counted in INSPIRE as of 09 Aug 2020



\bibitem{Faustov:2020thr}
  R.~N.~Faustov and V.~O.~Galkin,
  %``Semileptonic Decays of Heavy Baryons in the Relativistic Quark Model,''
  Particles {\bf 3}, no. 1, 208 (2020).
  % doi:10.3390/particles3010017
  %%CITATION = doi:10.3390/particles3010017;%%

\bibitem{Wang:2008sm}
  Y.~m.~Wang, Y.~Li and C.~D.~Lu,
  %``Rare Decays of Lambda(b) ---> Lambda + gamma and Lambda(b) ---> Lambda + l+ l- in the Light-cone Sum Rules,''
  Eur.\ Phys.\ J.\ C {\bf 59}, 861 (2009)
  % doi:10.1140/epjc/s10052-008-0846-5
  [arXiv:0804.0648 [hep-ph]].
  %%CITATION = doi:10.1140/epjc/s10052-008-0846-5;%%
  %77 citations counted in INSPIRE as of 11 Jul 2020


\end{thebibliography}
\end{document}